\documentclass[aps,prd,preprint,nofootinbib,11pt]{revtex4}
\usepackage{amsfonts}
\usepackage{mathrsfs}
\usepackage{graphicx}
\usepackage{amsmath}
\usepackage{amssymb}
\usepackage{subfigure}
\usepackage{epsfig}
\usepackage{graphicx}
\usepackage{ulem}
\usepackage{color}
\newcommand{\bqa}{\begin{eqnarray}}
\newcommand{\eqa}{\end{eqnarray}}
\newcommand{\beq}{\begin{equation}}
\newcommand{\eeq}{\end{equation}}
\allowdisplaybreaks[1]
\graphicspath{{fig/}{dia/}} \DeclareGraphicsExtensions{.eps}

\begin{document}
\title{\Large Mass spectra of $0^{--}$ and $0^{+-}$ hidden-heavy baryoniums\\[7mm]}

\author{Bing-Dong Wan$^{1,2,3}$\footnote{wanbd@lnnu.edu.cn} \vspace{+3pt}}

\affiliation{$^1$Department of Physics, Liaoning Normal University, Dalian 116029, China\\
$^2$ Center for Theoretical and Experimental High Energy Physics, Liaoning Normal University, Dalian 116029, China\\
$^3$ School of Fundamental Physics and Mathematical Sciences, Hangzhou Institute for Advanced Study, UCAS, Hangzhou 310024, China
}

\author{~\\~\\}

\begin{abstract}
\vspace{0.3cm}
In this work, the spectra of the prospective exotic hidden-charm and hidden-bottom baryonium, viz. the baryon-antibaryon states, with $J^{PC}=0^{--}$ and $0^{+-}$ are investigated in the framework of QCD sum rules. The non-perturbative contributions up to dimension 12 are taken into account. Numerical results indicate that there might exist 3 possible $0^{--}$ hidden-charm baryonium states with masses $(5.22\pm0.26)$, $(5.52\pm0.25)$, and $(5.46\pm0.24)$ GeV, and 5 possible $0^{+-}$ hidden-charm baryonium states with masses $(4.76\pm0.28)$, $(5.24\pm0.28)$, $(5.16\pm0.27)$, $(5.52\pm0.27)$, and $(5.69\pm0.27)$ GeV, respectively. The corresponding hidden-bottom partners are found lying in the range of $11.68-12.28$ GeV and $11.38-12.33$ GeV, respectively. The possible baryonium decay modes are analyzed, which are hopefully measurable in LHC experiments.
 \end{abstract}
 \maketitle
\newpage

\section{Introduction}

Since the observation of X(3872)~\cite{Choi:2003ue}, many novel hadronic states or candidates, denoted as $XYZ$ and $P_c$ states, have been observed in experiments. 
These novel hadronic states can hardly be well understood by conventional quark model~\cite{GellMann:1964nj,Zweig}, and exploring their properties and finding more possible states may greatly enrich the hadron family and our knowledge of the nature of quantum chromodynamics(QCD). With the more novel hadronic states discovering and the deepening of our understanding of strong interaction, it is reasonable to believe that the renaissance of hadron physics will come sooner or later, which attracts more and more interests from theorists and experimentalists. 

Among the novel hadronic states, special attention ought be paid to the states possess some exotic quantum numbers (like $0^{--}$, $0^{+-}$, $1^{-+}$, $2^{+-}$, and so on), since they cannot mix with the conventional ones. In the literature, many theoretical investigations on exotic hadronic states were made, including tetraquarks~\cite{Jiao:2009ra,LEE:2020eif,Shen:2010ky,Wang:2021lkg,General:2007bk,Huang:2016rro,Wan:2022xkx}, hybrid tates~\cite{Liu:2005rc,Chetyrkin:2000tj,Govaerts:1984bk,General:2006ed,Ishida:1991mx}, glueballs~\cite{Qiao:2014vva,Tang:2015twt,Cotanch:2006wv,Bellantuono:2015fia,Chen:2021cjr,Zhang:2021itx,Zhang:2022obn,Chen:2021bck,Ballon-Bayona:2017sxa}. On the other hand, exotic hadronic states have gradually been observed in experiment, like the most recently observed $\eta_1(1855)$~\cite{BESIII:2022riz,BESIII:2022iwi}. It is highly expected that more novel hadronic structures with exotic quantum numbers will emerge soon afterwards.

While the theoretical investigations on hexaquark with exotic quantum number are still rare.  Up to now, most of studies on hexaquark are focus on conventional quantum numbers, such as the deuteron. Deuteron created at the beginning of Universe and its stability is responsible for the production of other elements, is a well-established dibaryon molecular state composed of a proton and a neutron, withe $J^P=1^+$ and binding energy $E_B=2.225$ MeV~\cite{Weinberg:1962hj}. On the other hand, the baryon-antibaryon systems (baryonium) is another special class of heaxquark configuration. 
 Actually, the history of the investigation on baryon-antibaryon system can date back to the proposal by Fermi and Yang that $\pi$-mesons may be composite particles formed by a nucleon-antinucleon pair in 1940s~\cite{Fermi:1949voc}, and their scenario was later on replaced by the quark model. Entering the new millennium, the heavy baryonium were proposed and employed to explain the extraordinary nature of $Y(4260)$ \cite{Qiao:2005av, Qiao:2007ce} and other charmonium-like states observed in experiments. Later on, more investigations on baryonium are performed from various aspects \cite{Chen:2011cta, Chen:2013sba,Wan:2019ake,Chen:2016ymy,Liu:2007tj,Wang:2021qmn,Wan:2022uie,Liu:2021gva,Wan:2021vny}. Especially, our previous work~\cite{Wan:2021vny} of the investigation of light baryonium states predicted that the mass of the $\Lambda-\bar{\Lambda}$ baryonium state with $J^{PC}$ of $1^{--}$ is about $2.34$ GeV, which is consistent with the experimental observation of the structure in $\Lambda\bar{\Lambda}$ mass spectra with mass of $(2356\pm24)$ MeV with quantum number of $1^{--}$ by BESIII~\cite{BESIII:2022tvj}.

In this work, we investigate the hidden-heavy baryonium states with the exotic quantum numbers of $J^{PC}=0^{--}$ and  $0^{+-}$ in the framework of QCD sum rules (QCDSR). In past 40 years, the QCDSR technique~\cite{Shifman} as a model independent approach has achieved a lot in the study of hadron spectroscopy. To establish the QCDSR, one need to construct the proper interpolating currents corresponding to hadrons of interest, which possesses the foremost information of the concerned hadrons, like quantum numbers and structure components. With the interpolating currents, the two-point correlation function can be readily established and by equating the operator product expansion (OPE) side and the phenomenological side of the two-point correlation, we can then obtain the mass of the hadron.

The rest of the paper is arranged as follows. After the introduction, a brief interpretation of QCD sum rules and some primary formulas in our calculation are presented in Sec. \ref{Formalism}. The numerical analysis and results are given in Sec. \ref{Numerical}. The possible decay modes of the exotic baryonium states are analyzed in Sec. \ref{decay}. The last part is left for a brief summary.

\section{Formalism}\label{Formalism}

For the $0^{--}$ baryonium state, the interpolating currents in molecular configuration can be constructed as:
\begin{eqnarray}\label{current_0--}
j_{0^{--}}^A(x)&=&\frac{\epsilon_{a b c}\epsilon_{d e f}}{\sqrt{2}} {\Big\{}[\bar{Q}_d(x) Q_c(x)][q^T_a(x) C q^\prime_b(x)][\bar{q}_e (x)\gamma_5 C \bar{q}^{\prime T}_f(x)] \nonumber\\
&-&[\bar{Q}_d(x) Q_c(x)][q^T_a(x) C \gamma_5 q^\prime_b(x)][\bar{q}_e (x) C \bar{q}^{\prime T}_f(x)] {\Big\}}\;,\label{Ja0--}\\
j_{0^{--}}^B(x)&=&\frac{\epsilon_{a b c}\epsilon_{d e f}}{\sqrt{2}}{\Big\{}[\bar{Q}_d(x) Q_c(x)][q^T_a(x) C \gamma_\mu q^\prime_b(x)][\bar{q}_e(x) \gamma_\mu\gamma_5 C \bar{q}^{\prime T}_f(x)] \nonumber\\
&-&[\bar{Q}_d(x) Q_c(x)][q^T_a(x) C \gamma_\mu\gamma_5 q^\prime_b(x)][\bar{q}_e (x) \gamma_\mu C \bar{q}^{\prime T}_f(x)] {\Big\}}\;,\label{Jb0--}\\
j_{0^{--}}^C(x)&=& \frac{\epsilon_{a b c}\epsilon_{d e f}}{\sqrt{2}} {\Big\{}[\bar{Q}_d(x) \gamma_\mu Q_c(x)][q^T_a(x) C \gamma_\mu q^\prime_b(x)][\bar{q}_e (x) \gamma_5 C \bar{q}^{\prime T}_f (x)]\nonumber\\
&+&[\bar{Q}_d(x) \gamma_\mu Q_c(x)][q^T_a(x) C \gamma_5 q^\prime_b(x)][\bar{q}_e (x) \gamma_\mu C \bar{q}^{\prime T}_f (x)] {\Big\}}\;,\label{Jc0--}\\
j_{0^{--}}^D(x)&=&\frac{\epsilon_{a b c}\epsilon_{d e f}}{\sqrt{2}} {\Big\{}[\bar{Q}_d(x) \gamma_\mu Q_c(x)][q^T_a(x) C \gamma_\mu\gamma_5 q^\prime_b(x)][\bar{q}_e (x) C \bar{q}^{\prime T}_f (x)]\nonumber\\
&+&[\bar{Q}_d(x) \gamma_\mu Q_c(x)][q^T_a(x) C q^\prime_b(x)][\bar{q}_e (x) \gamma_\mu\gamma_5 C \bar{q}^{\prime T}_f (x)] {\Big\}}\;, \label{Jd0--}\\
j_{0^{--}}^E(x)&=& \frac{\epsilon_{a b c}\epsilon_{d e f}}{\sqrt{2}} {\Big\{}[\bar{Q}_d(x) \gamma_\mu\gamma_5 Q_c(x)][q^T_a(x) C \gamma_\mu\gamma_5 q^\prime_b(x)][\bar{q}_e (x) \gamma_5 C \bar{q}^{\prime T}_f (x)]\nonumber\\
&-&[\bar{Q}_d(x) \gamma_\mu\gamma_5 Q_c(x)][q^T_a(x) C \gamma_5 q^\prime_b(x)][\bar{q}_e (x) \gamma_\mu\gamma_5 C \bar{q}^{\prime T}_f (x)] {\Big\}}\;,\label{Je0--}\\
j_{0^{--}}^F(x)&=&\frac{\epsilon_{a b c}\epsilon_{d e f}}{\sqrt{2}}{\Big\{} [\bar{Q}_d(x)  \gamma_\mu\gamma_5 Q_c(x)][q^T_a(x) C \gamma_\mu q^\prime_b(x)][\bar{q}_e (x) C \bar{q}^{\prime T}_f (x)]\nonumber\\
&-& [\bar{Q}_d(x)  \gamma_\mu\gamma_5 Q_c(x)][q^T_a(x) C q^\prime_b(x)][\bar{q}_e (x) \gamma_\mu C \bar{q}^{\prime T}_f (x)] {\Big\}}\;. \label{Jf0--}
\end{eqnarray}
The interpolating currents for $0^{+-}$ in molecular configuration are found to be in forms:
\begin{eqnarray}\label{current_0+-}
j_{0^{+-}}^A(x)&=&\frac{\epsilon_{a b c}\epsilon_{d e f}}{\sqrt{2}} {\Big\{}[\bar{Q}_d(x) \gamma_5 Q_c(x)][q^T_a(x) C q^\prime_b(x)][\bar{q}_e (x)\gamma_5 C \bar{q}^{\prime T}_f(x)] \nonumber\\
&-&[\bar{Q}_d(x) \gamma_5 Q_c(x)][q^T_a(x) C \gamma_5 q^\prime_b(x)][\bar{q}_e (x) C \bar{q}^{\prime T}_f(x)] {\Big\}}\;,\label{Ja0+-}\\
j_{0^{+-}}^B(x)&=&\frac{\epsilon_{a b c}\epsilon_{d e f}}{\sqrt{2}}{\Big\{}[\bar{Q}_d(x)\gamma_5 Q_c(x)][q^T_a(x) C \gamma_\mu q^\prime_b(x)][\bar{q}_e(x) \gamma_\mu\gamma_5 C \bar{q}^{\prime T}_f(x)] \nonumber\\
&-&[\bar{Q}_d(x) \gamma_5 Q_c(x)][q^T_a(x) C \gamma_\mu\gamma_5 q^\prime_b(x)][\bar{q}_e (x) \gamma_\mu C \bar{q}^{\prime T}_f(x)] {\Big\}}\;,\label{Jb0+-}\\
j_{0^{+-}}^C(x)&=& \frac{\epsilon_{a b c}\epsilon_{d e f}}{\sqrt{2}} {\Big\{}[\bar{Q}_d(x) \gamma_\mu Q_c(x)][q^T_a(x) C \gamma_\mu q^\prime_b(x)][\bar{q}_e (x) C \bar{q}^{\prime T}_f (x)]\nonumber\\
&+&[\bar{Q}_d(x) \gamma_\mu Q_c(x)][q^T_a(x) C q^\prime_b(x)][\bar{q}_e (x) \gamma_\mu C \bar{q}^{\prime T}_f (x)] {\Big\}}\;,\label{Jc0+-}\\
j_{0^{+-}}^D(x)&=&\frac{\epsilon_{a b c}\epsilon_{d e f}}{\sqrt{2}} {\Big\{}[\bar{Q}_d(x) \gamma_\mu Q_c(x)][q^T_a(x) C \gamma_\mu\gamma_5 q^\prime_b(x)][\bar{q}_e (x) \gamma_5 C \bar{q}^{\prime T}_f (x)]\nonumber\\
&+&[\bar{Q}_d(x) \gamma_\mu Q_c(x)][q^T_a(x) C \gamma_5 q^\prime_b(x)][\bar{q}_e (x) \gamma_\mu\gamma_5 C \bar{q}^{\prime T}_f (x)] {\Big\}}\;, \label{Jd0+-}\\
j_{0^{+-}}^E(x)&=& \frac{\epsilon_{a b c}\epsilon_{d e f}}{\sqrt{2}} {\Big\{}[\bar{Q}_d(x) \gamma_\mu\gamma_5 Q_c(x)][q^T_a(x) C \gamma_\mu q^\prime_b(x)][\bar{q}_e (x) \gamma_5 C \bar{q}^{\prime T}_f (x)]\nonumber\\
&-&[\bar{Q}_d(x) \gamma_\mu\gamma_5 Q_c(x)][q^T_a(x) C \gamma_5 q^\prime_b(x)][\bar{q}_e (x) \gamma_\mu C \bar{q}^{\prime T}_f (x)] {\Big\}}\;,\label{Je0+-}\\
j_{0^{+-}}^F(x)&=&\frac{\epsilon_{a b c}\epsilon_{d e f}}{\sqrt{2}}{\Big\{} [\bar{Q}_d(x)  \gamma_\mu\gamma_5 Q_c(x)][q^T_a(x) C \gamma_\mu\gamma_5 q^\prime_b(x)][\bar{q}_e (x) C \bar{q}^{\prime T}_f (x)]\nonumber\\
&-& [\bar{Q}_d(x)  \gamma_\mu\gamma_5 Q_c(x)][q^T_a(x) C q^\prime_b(x)][\bar{q}_e (x) \gamma_\mu\gamma_5 C \bar{q}^{\prime T}_f (x)] {\Big\}}\;. \label{Jf0+-}
\end{eqnarray}

Here, the subscripts $a\cdots f$ are color indices, $q$ and $q^\prime$ stand for light quark $u$ and $d$, respectively, $Q$ represents the heavy quark $c$ or $b$, and $C$ is the charge conjugation matrix.

With the currents (\ref{Ja0--})-(\ref{Jf0+-}), the two-point correlation function can be readily established, i.e.,
\begin{eqnarray}
\Pi_{J^{PC}}^k(q^2) &=& i \int d^4 x e^{i q \cdot x} \langle 0 | T \{ j_{J^{PC}}^k(x),\;  j_{J^{PC}}^k (0)^\dagger \} |0 \rangle \;,
\end{eqnarray}
where $ |0 \rangle$ denotes the physical vacuum, and $k$ runs from $A$ to $F$. 

On the OPE side, correlation function $\Pi(q^2)$ can be expressed by the dispersion relation as
 \begin{eqnarray}
\Pi^{OPE,\;k}_{J^{PC}} (q^2) = \int_{s_{min}}^{\infty} d s\frac{\rho^{OPE,\;k}_{J^{PC}}(s)}{s - q^2}\; .
\label{OPE-hadron}
\end{eqnarray}
Here, $s_{min}$ is the kinematic limit, which usually corresponds to the square of the sum of current-quark masses of the hadron \cite{Albuquerque:2013ija}, $\rho^{OPE,\;k}_{J^{PC}}(s) = \text{Im} [\Pi^{OPE,\;k}_{J^{PC}}(s)] / \pi$, and
\begin{eqnarray}
\rho^{OPE}(s) & = & \rho^{pert}(s)  +\rho^{\langle G^2 \rangle}(s) + \rho^{\langle \bar{q} G q \rangle}(s)
+ \rho^{\langle \bar{q} q \rangle^2}(s)
+\rho^{\langle G^3 \rangle}(s)\nonumber\\
&+& \rho^{\langle \bar{q} q \rangle\langle \bar{q} G q \rangle}(s)
+ \rho^{\langle \bar{q} q \rangle\langle \bar{q} G q \rangle}(s) 
+ \rho^{\langle \bar{q} G q \rangle^2}(s) \;. \label{rho-OPE}
\end{eqnarray}
 The Feynman diagrams corresponding to each term of Eq. (\ref{rho-OPE}) are schematically shown in Fig. \ref{feyndiag}. The analytical expressions of $\rho^{OPE,\;k}_{J^{PC}}(s)$ can be calculated and given in appendix \ref{ana_exp}.

\begin{figure}
\includegraphics[width=8.8cm]{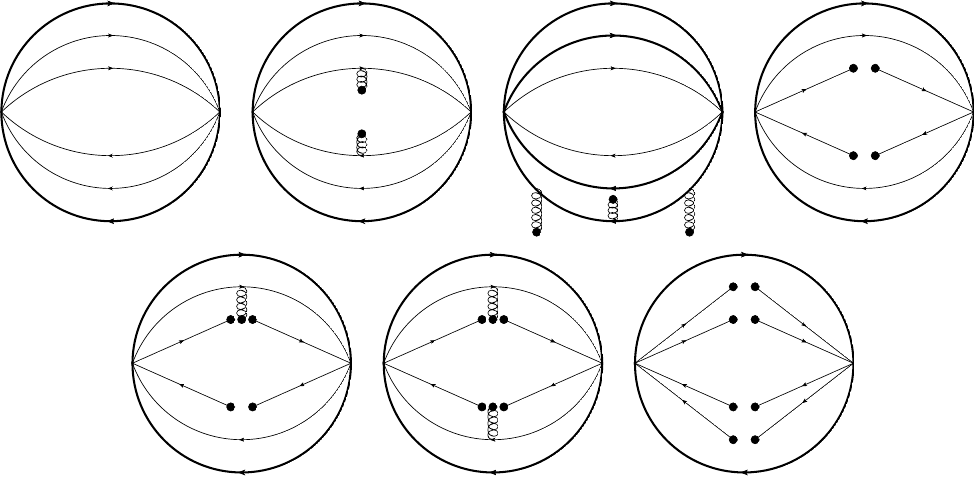}
\caption{The typical Feynman diagrams related to the correlation function, where the thick solid line represents the heavy quark, the thin solid line stands for the light quark, and the spiral line denotes the gluon. There is no heavy quark consendsate due to the large heavy quark mass.} \label{feyndiag}
\end{figure}

On the phenomenological side, the correlation function $\Pi(q^2)$ can be expressed as a dispersion integral over the physical regime after isolating the ground state contribution from the baryonium state, i.e.,
 \begin{eqnarray}
\Pi^{phen,\;k}_{J^{PC}}(q^2) & = & \frac{{\lambda^k_{J^{PC}}}^2}{{M^k_{J^{PC}}}^2 - q^2} + \frac{1}{\pi} \int_{s_0}^\infty d s \frac{\rho^k_{J^{PC}}(s)}{s - q^2} \; , \label{hadron}
\end{eqnarray}
where $M$ denotes the mass of baryonium state, $\lambda$ is the coupling constant, and $\rho(s)$ is the spectral density that contains the contributions from higher excited states and the continuum states above the threshold $s_0$.

Performing Borel transform on Eqs.~(\ref{OPE-hadron}) and (\ref{hadron}), and matching the OPE side with the phenomenological side of the correlation function $\Pi(q^2)$, one can finally obtain the mass of the tetraquark state,
\begin{eqnarray}
M^k_{J^{PC}}(s_0, M_B^2) = \sqrt{- \frac{L^k_{J^{PC},\;1}(s_0, M_B^2)}{L^k_{J^{PC},\;0}(s_0, M_B^2)}} \; . \label{mass-Eq}
\end{eqnarray}
Here $L_0$ and $L_1$ are respectively defined as
\begin{eqnarray}
L^k_{J^{PC},\;0}(s_0, M_B^2) =  \int_{s_{min}}^{s_0} d s \; \rho^{OPE,\; k}_{J^{PC}}(s) e^{-
s / M_B^2}   \;,  \label{L0}
\end{eqnarray}
and
\begin{eqnarray}
L^k_{J^{PC},\;1}(s_0, M_B^2) =
\frac{\partial}{\partial{\frac{1}{M_B^2}}}{L^k_{J^{PC},\;0}(s_0, M_B^2)} \; .
\end{eqnarray}

\section{Numerical analysis}\label{Numerical}

In performing the numerical calculation, the broadly accepted inputs are taken \cite{Albuquerque:2013ija,Matheus:2006xi, Cui:2011fj, Narison:2002pw,P.Col,Tang:2019nwv}:
\begin{eqnarray}
\begin{aligned}
&m_c(m_c)=\overline{m}_c=(1.275\pm0.025)\; \text{GeV}\;,&&m_b(m_b)=\overline{m}_b=(4.18\pm0.03)\; \text{GeV}   \; ,\\
& \langle \bar{q} q \rangle = - (0.24 \pm 0.01)^3 \; \text{GeV}^3 \; ,& & \langle g_s^2 G^2 \rangle = (0.88\pm0.25) \; \text{GeV}^4 \; , \\
& \langle \bar{q} g_s \sigma \cdot G q \rangle = m_0^2 \langle\bar{q} q \rangle; , & & \langle g_s^3 G^3 \rangle = (0.045 \pm 0.013) \;\text{GeV}^6 \; ,\\
& m_0^2 = (0.8 \pm 0.1) \; \text{GeV}^2\; .
\end{aligned}
\end{eqnarray}
Here, $\overline{m}_c$ and $\overline{m}_b$ represent heavy-quark running masses in $\overline{\text{MS}}$ scheme. For light quark, the chiral quark limit mass $m_q=0$ is adopted. Furthermore, as the discussion in Ref. \cite{Albuquerque:2016znh}, we cannot discard the on-shell quark masses which are $m_c\approx 1.5$ GeV and $m_b\approx 4.7$ GeV, so the on-shell quark masses are also taken into consideration in our calculation.

Moreover, two additional parameters $s_0$ and $M_B^2$ introduced in establishing the sum rules need to be exacted. We can fix them in light of the so-called standard procedures by fulfilling the following two criteria \cite{Shifman,Reinders:1984sr, P.Col}. One asks for the convergence of the OPE, which is to compare relative contribution of higher dimension condensate to the total contribution on the OPE side, and then a reliable region for $M_B^2$ will be chosen to retain the convergence. The other one requires that the pole contribution (PC) is more than $15\%$ of the total for hexaquark states~\cite{Wan:2019ake,Wan:2022uie,Wan:2021vny}. Mathematically, the two criteria can be formulated as:
\begin{eqnarray}
  R^{OPE}= \left| \frac{L_{J^{PC}\;,0}^{k\;,dim=12}(s_0, M_B^2)}{L_{J^{PC}\;,0}^k(s_0, M_B^2)}\right|\, ,
\end{eqnarray}
\begin{eqnarray}
  R^{PC} = \frac{L_{J^{PC}\;,0}^k(s_0, M_B^2)}{L_{J^{PC}\;,0}^k(\infty, M_B^2)} \; . \label{RatioPC}
\end{eqnarray}

To find a proper value for continuum threshold $s_0$, a similar analysis as in Refs. \cite{Qiao:2013dda,Tang:2016pcf,Qiao:2013raa,Wan:2020oxt,Wan:2020fsk} is performed. Therein, one needs to find the proper value, which has an optimal window for the mass curve of the baryonium state. Within this window, the physical quantity, that is the mass of the baryonium state, should be independent of the Borel parameter $M_B^2$ as much as possible. In practice, we will vary $\sqrt{s_0}$ by $0.2$ GeV to obtain the lower and upper bounds, and hence the uncertainties of $\sqrt{s_0}$.

\begin{figure}
\includegraphics[width=6.8cm]{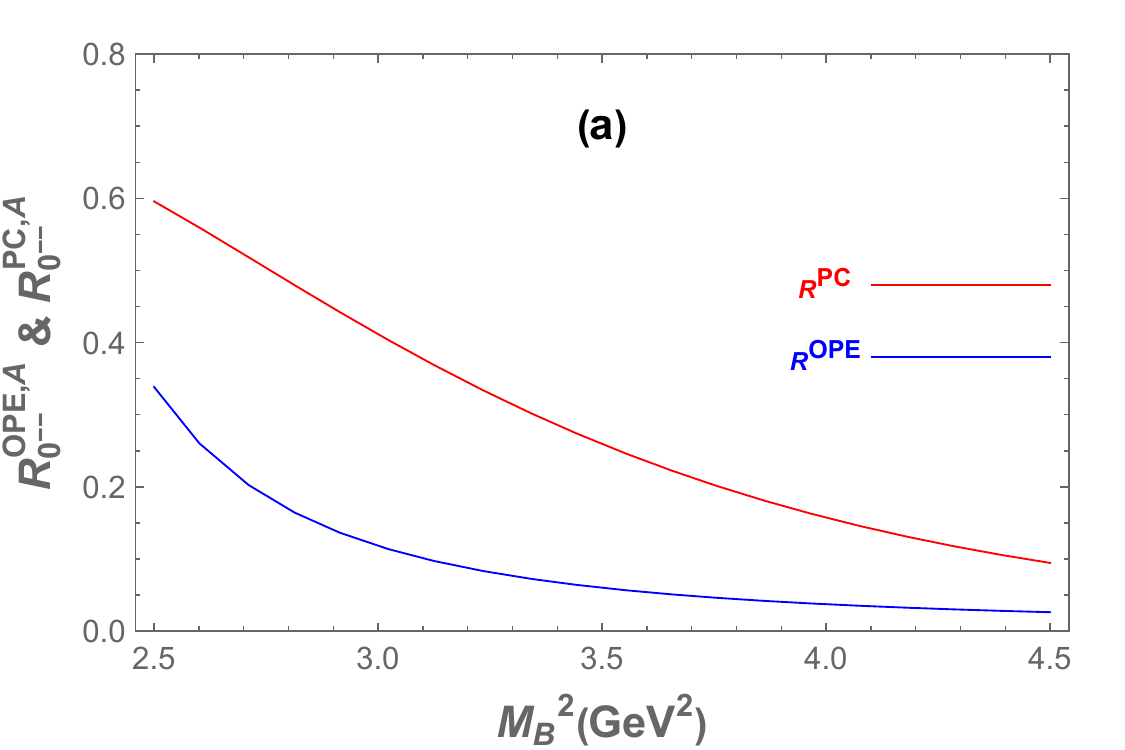}
\includegraphics[width=6.8cm]{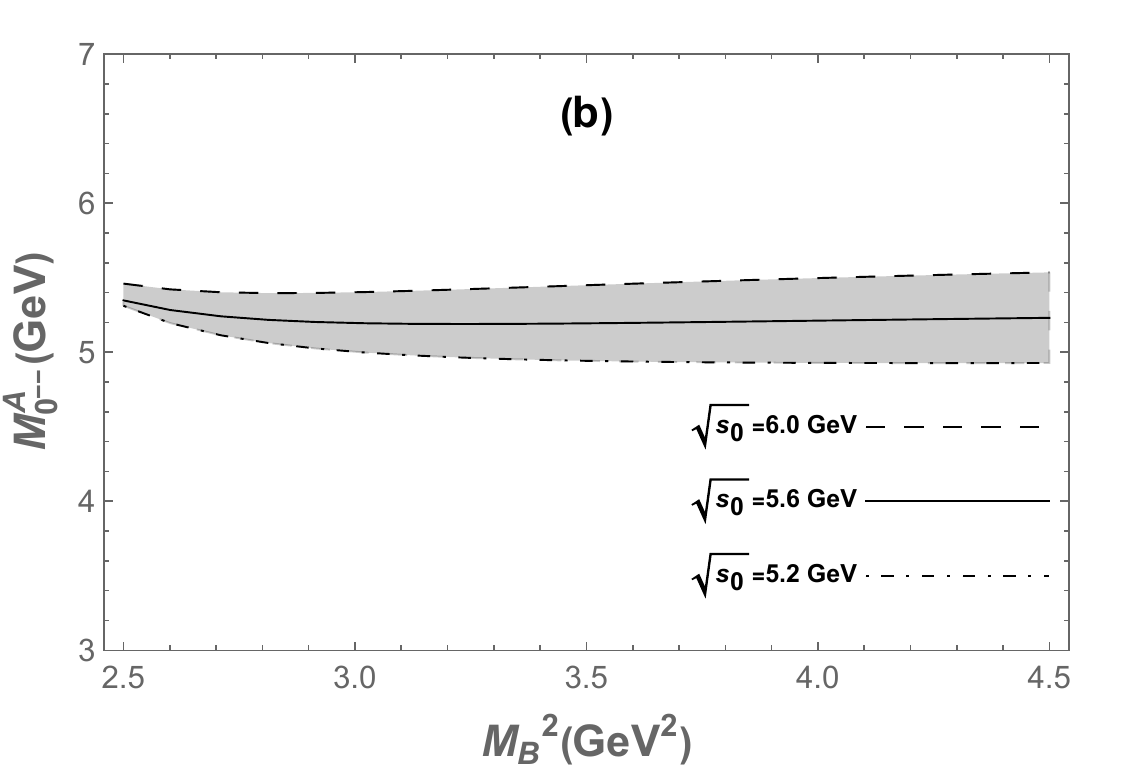}
\caption{ (a) The ratios of ${R_{A}^{OPE}}$ and ${R_{A}^{PC}}$ as functions of the Borel parameter $M_B^2$ for different values of $\sqrt{s_0}$, where blue lines represent ${R_{A}^{OPE}}$ and red lines denote ${R_{A}^{PC}}$ . (b) The mass $M^{A}$ as a function of the Borel parameter $M_B^2$ for different values of $\sqrt{s_0}$.} \label{figA0--}
\end{figure}

With the above preparation the mass spectrum of exotic baryonium states can be numerically evaluated. For the $0^{--}$ hidden-charm exotic baryonium state in Eq.~(\ref{Ja0--}),  the ratios $R^{OPE\;,A}_{0^{--}}$ and $R^{PC\;,A}_{0^{--}}$ are presented as functions of Borel parameter $M_B^2$ in Fig. \ref{figA0--}(a) with different values of $\sqrt{s_0}$, i.e., $5.2$, $5.6$ and $6.0$ GeV. The reliant relations of $M^{A}_{0^{--}}$ on parameter $M_B^2$ are displayed in Fig. \ref{figA0--}(b). The optimal Borel window is found in range $3.1 \le M_B^2 \le 4.1\; \text{GeV}^2$, and the mass $M^{A}_{0^{--}}$ can then be obtained:
\begin{eqnarray}
M^{A}_{0^{--}} &=& (5.22\pm 0.26)\; \text{GeV}.\label{m1}
\end{eqnarray}

\begin{figure}[h]
\includegraphics[width=6.8cm]{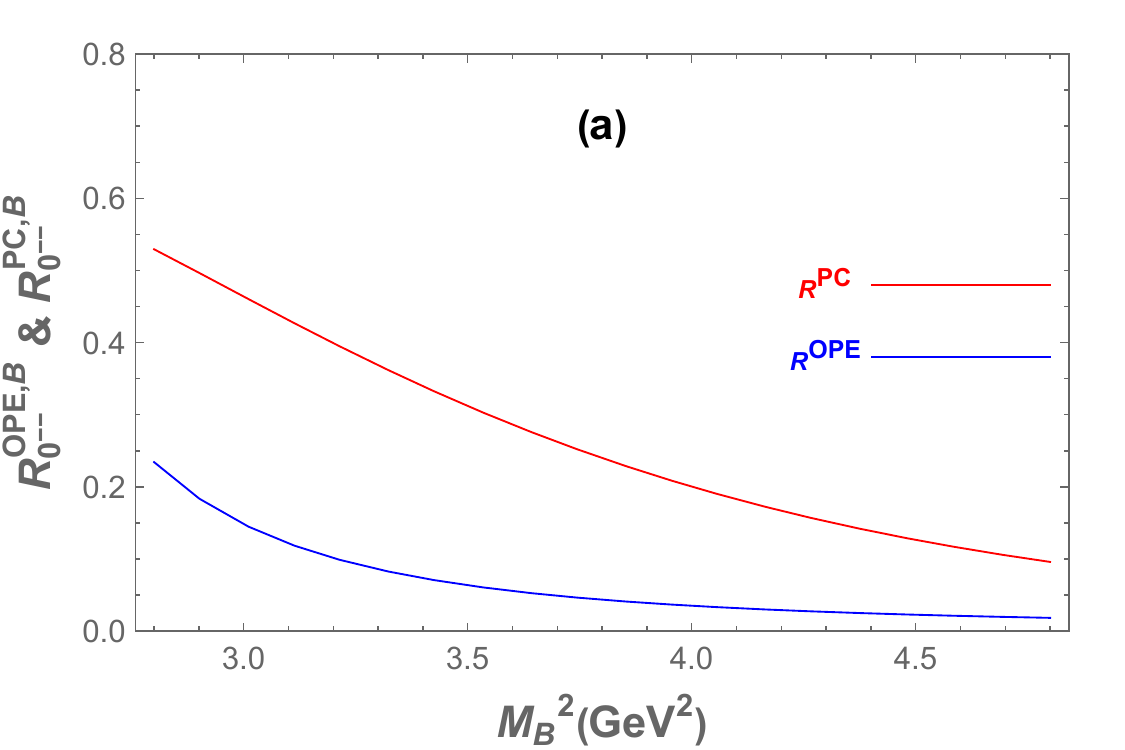}
\includegraphics[width=6.8cm]{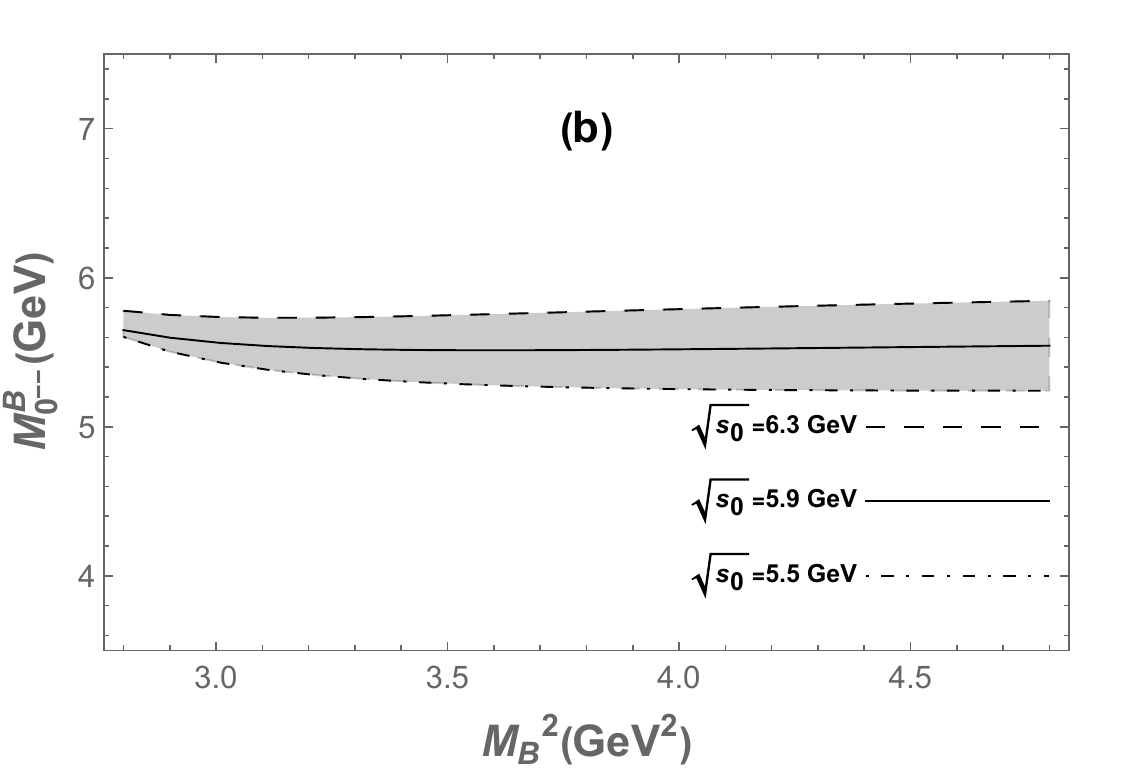}
\caption{The same caption as in Fig \ref{figA0--}, but for the current in Eq.~(\ref{Jb0--}).} \label{figB0--}
\end{figure}

For the $0^{--}$ hidden-charm exotic baryonium state in Eq.~(\ref{Jb0--}), the ratios $R^{OPE\;,B}_{0^{--}}$ and $R^{PC\;,B}_{0^{--}}$ are presented as functions of Borel parameter $M_B^2$ in Fig. \ref{figB0--}(a) with different values of $\sqrt{s_0}$, i.e., $5.5$, $5.9$ and $6.3$ GeV. The reliant relations of $M^{B}_{0^{--}}$ on parameter $M_B^2$ are displayed in Fig. \ref{figB0--}(b). The optimal Borel window is found in range $3.2 \le M_B^2 \le 4.3\; \text{GeV}^2$, and the mass $M^{B}_{0^{--}}$ can then be obtained:
\begin{eqnarray}
M^{B}_{0^{--}} &=& (5.52\pm 0.25)\; \text{GeV}.\label{m2}
\end{eqnarray}

\begin{figure}[h]
\includegraphics[width=6.8cm]{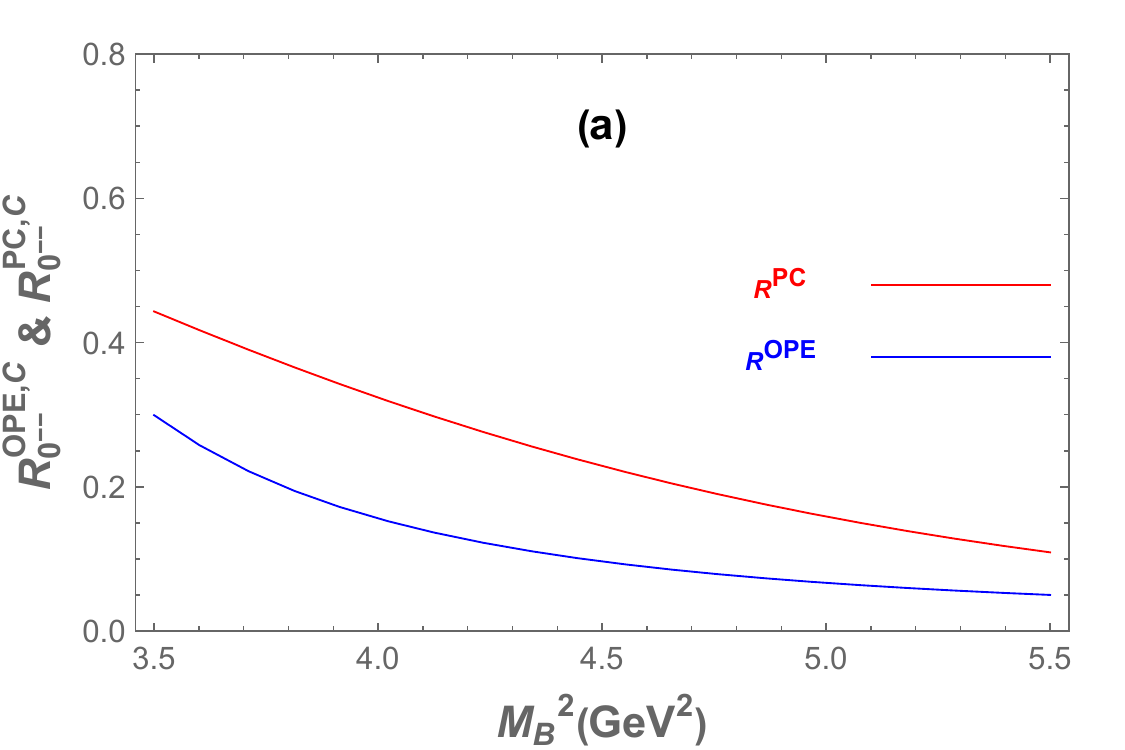}
\includegraphics[width=6.8cm]{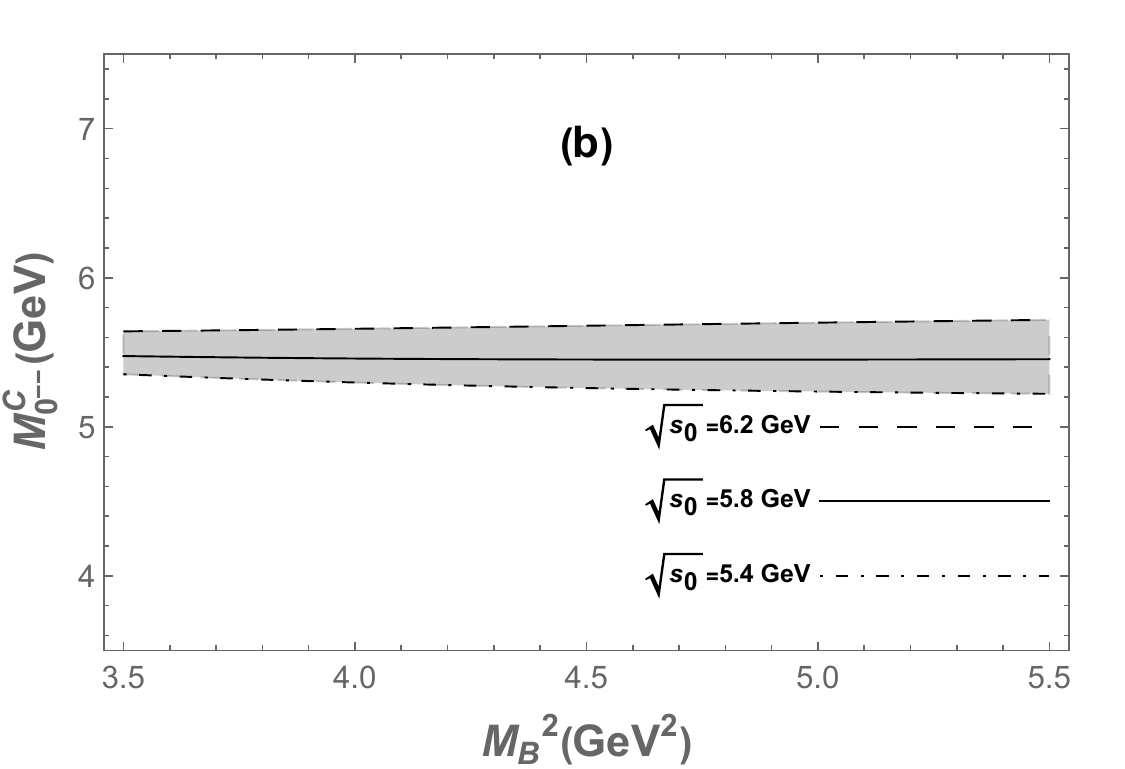}
\caption{The same caption as in Fig \ref{figA0--}, but for the current in Eq.~(\ref{Jc0--}).} \label{figC0--}
\end{figure}

For the $0^{--}$ hidden-charm exotic baryonium state in Eq.~(\ref{Jc0--}), the ratios $R^{OPE\;,C}_{0^{--}}$ and $R^{PC\;,C}_{0^{--}}$ are presented as functions of Borel parameter $M_B^2$ in Fig. \ref{figC0--}(a) with different values of $\sqrt{s_0}$, i.e., $5.4$, $5.8$ and $6.2$ GeV. The reliant relations of $M^{C}_{0^{--}}$ on parameter $M_B^2$ are displayed in Fig. \ref{figC0--}(b). The optimal Borel window is found in range $4.1 \le M_B^2 \le 5.1\; \text{GeV}^2$, and the mass $M^{C}_{0^{--}}$ can then be obtained:
\begin{eqnarray}
M^{C}_{0^{--}} &=& (5.46\pm 0.24)\; \text{GeV}.\label{m3}
\end{eqnarray}

\begin{figure}[h]
\includegraphics[width=6.8cm]{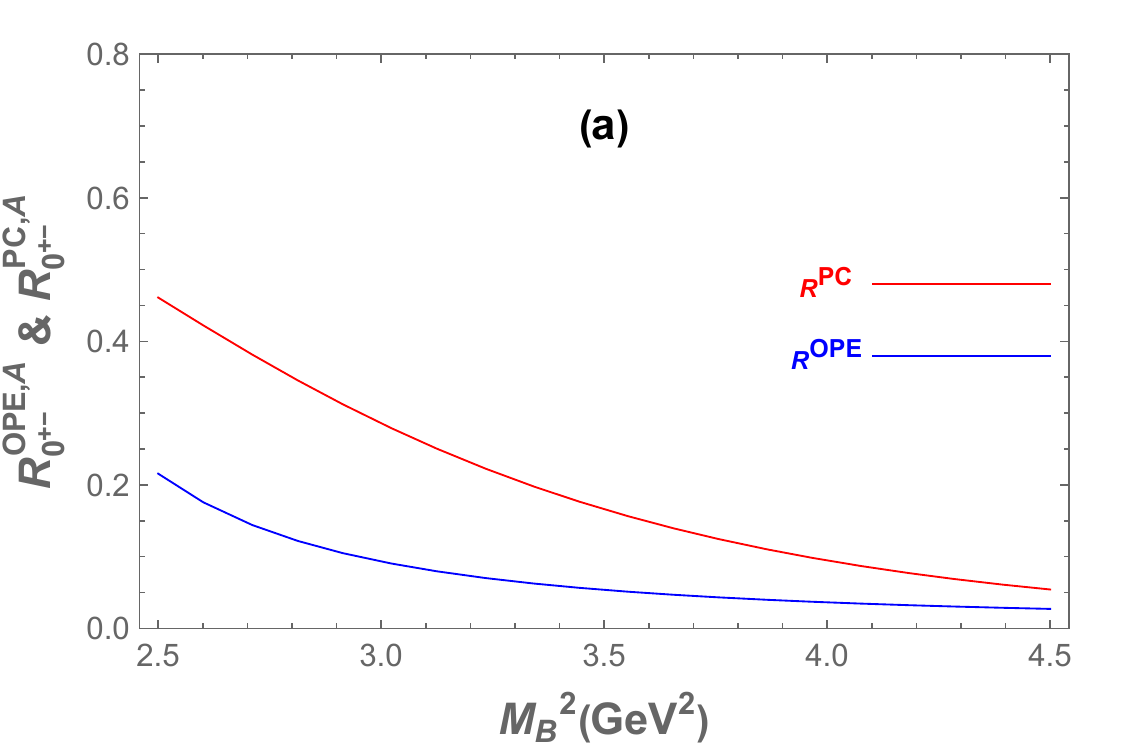}
\includegraphics[width=6.8cm]{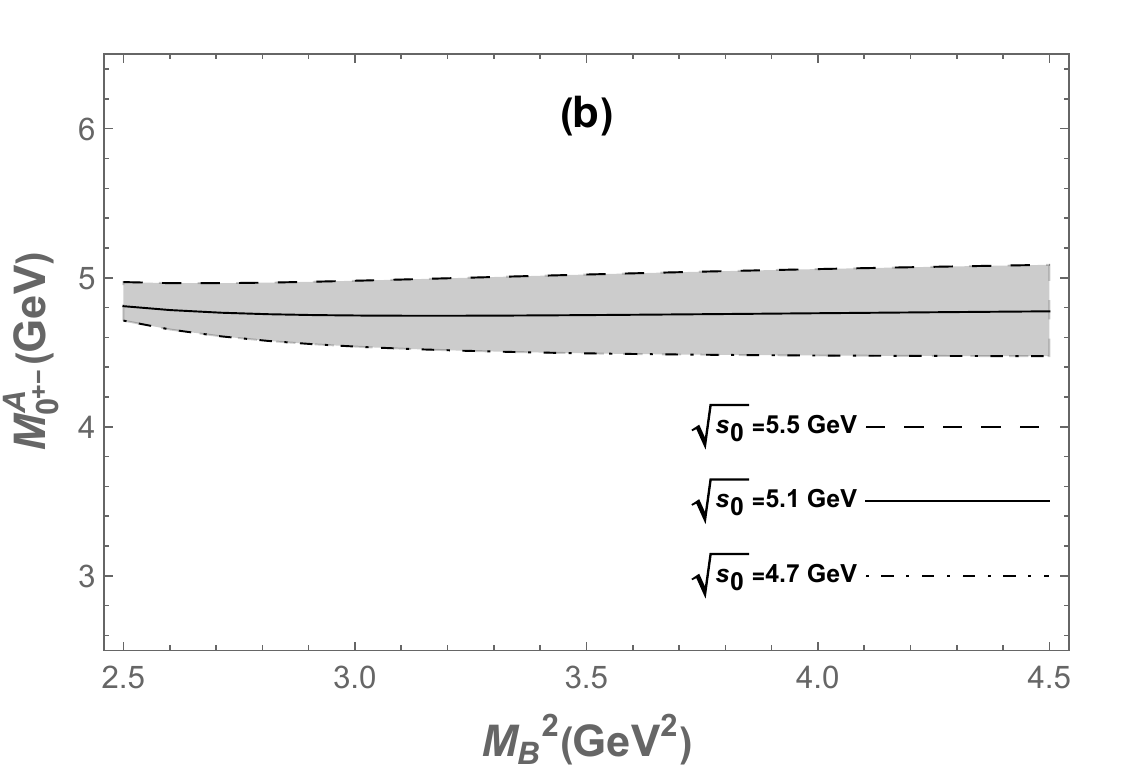}
\caption{The same caption as in Fig \ref{figA0--}, but for the current  in Eq.~(\ref{Ja0+-}).} \label{figA0+-}
\end{figure}

For the $0^{+-}$ hidden-charm exotic baryonium state in Eq.~(\ref{Ja0+-}), the ratios $R^{OPE\;,A}_{0^{+-}}$ and $R^{PC\;,A}_{0^{+-}}$ are presented as functions of Borel parameter $M_B^2$ in Fig. \ref{figA0+-}(a) with different values of $\sqrt{s_0}$, i.e., $4.7$, $5.1$ and $5.5$ GeV. The reliant relations of $M^{A}_{0^{+-}}$ on parameter $M_B^2$ are displayed in Fig. \ref{figA0+-}(b). The optimal Borel window is found in range $2.8 \le M_B^2 \le 3.8\; \text{GeV}^2$, and the mass $M^{A}_{0^{+-}}$ can then be obtained:
\begin{eqnarray}
M^{A}_{0^{+-}} &=& (4.76\pm 0.26)\; \text{GeV}.\label{m4}
\end{eqnarray}

\begin{figure}[h]
\includegraphics[width=6.8cm]{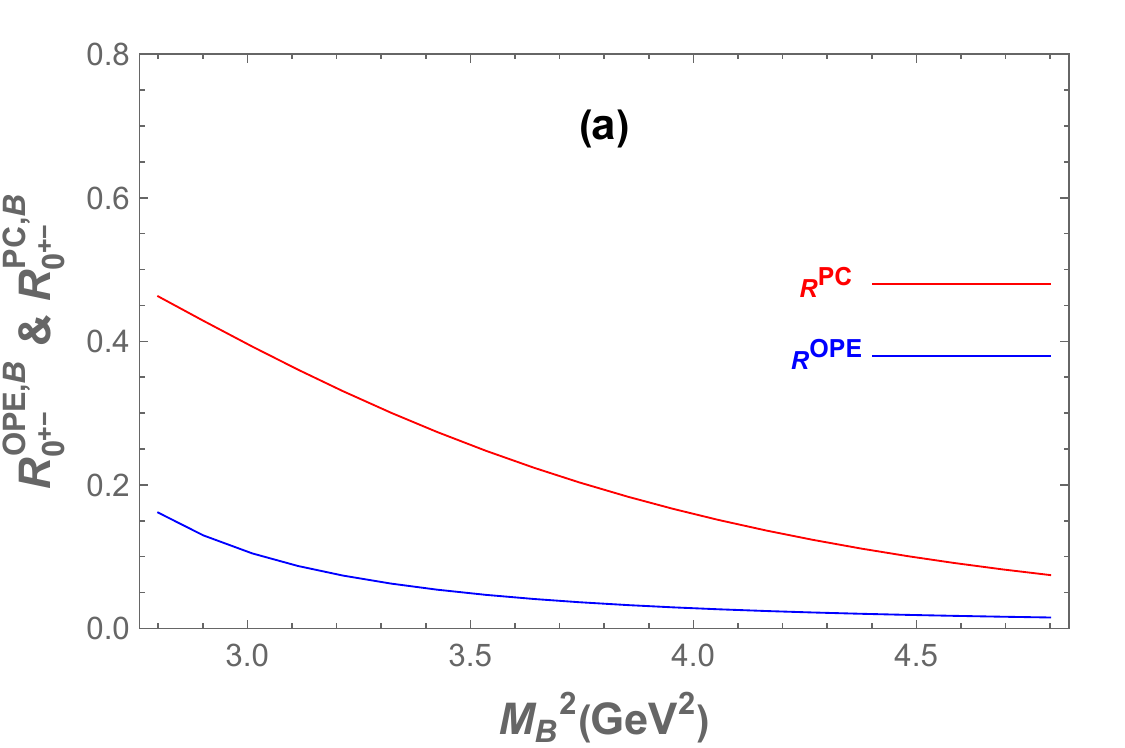}
\includegraphics[width=6.8cm]{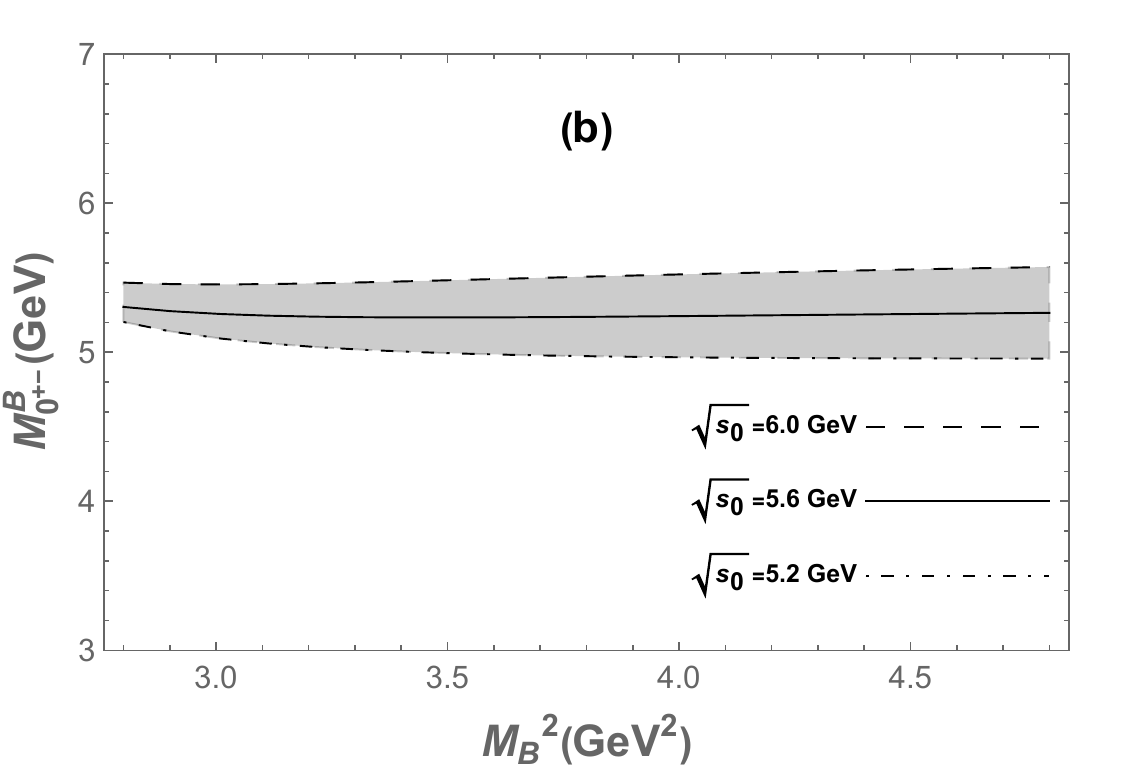}
\caption{The same caption as in Fig \ref{figA0--}, but for the current  in Eq.~(\ref{Jb0+-}).} \label{figB0+-}
\end{figure}

For the $0^{+-}$ hidden-charm exotic baryonium state in Eq.~(\ref{Jb0+-}), the ratios $R^{OPE\;,B}_{0^{+-}}$ and $R^{PC\;,B}_{0^{+-}}$ are presented as functions of Borel parameter $M_B^2$ in Fig. \ref{figB0+-}(a) with different values of $\sqrt{s_0}$, i.e., $5.2$, $5.6$ and $5.9$ GeV. The reliant relations of $M^{B}_{0^{+-}}$ on parameter $M_B^2$ are displayed in Fig. \ref{figB0+-}(b). The optimal Borel window is found in range $3.0 \le M_B^2 \le 4.1\; \text{GeV}^2$, and the mass $M^{B}_{0^{+-}}$ can then be obtained:
\begin{eqnarray}
M^{B}_{0^{+-}} &=& (5.24\pm 0.28)\; \text{GeV}.\label{m5}
\end{eqnarray}

\begin{figure}[h]
\includegraphics[width=6.8cm]{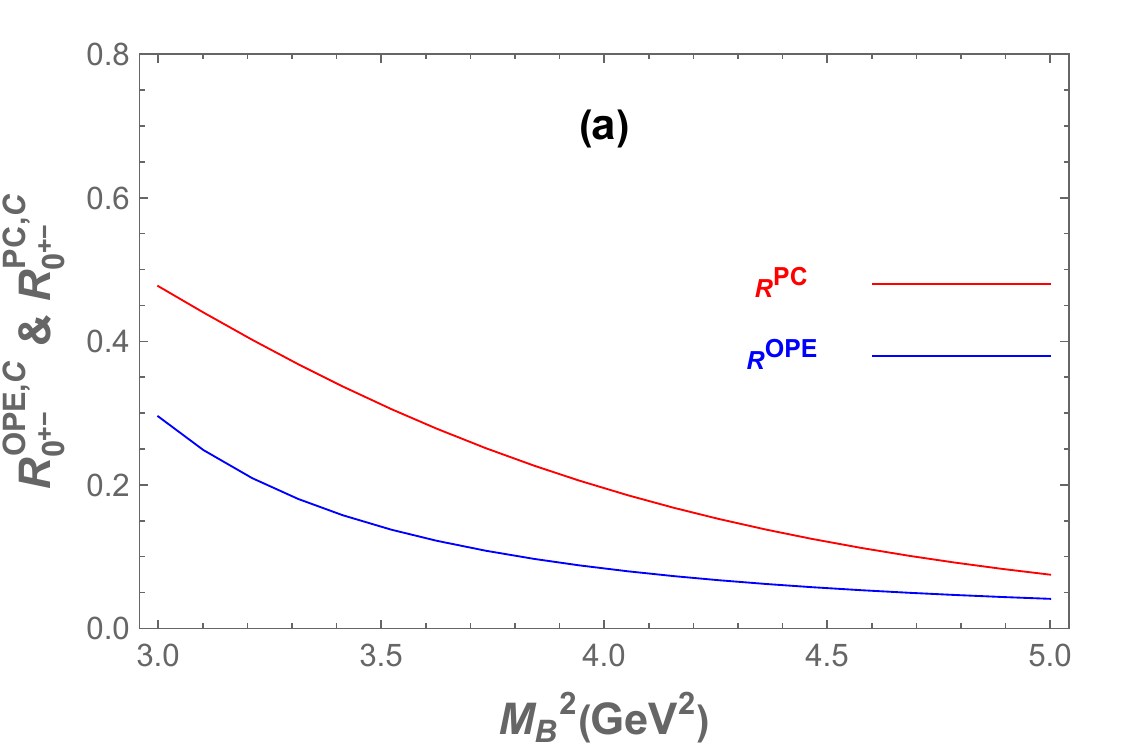}
\includegraphics[width=6.8cm]{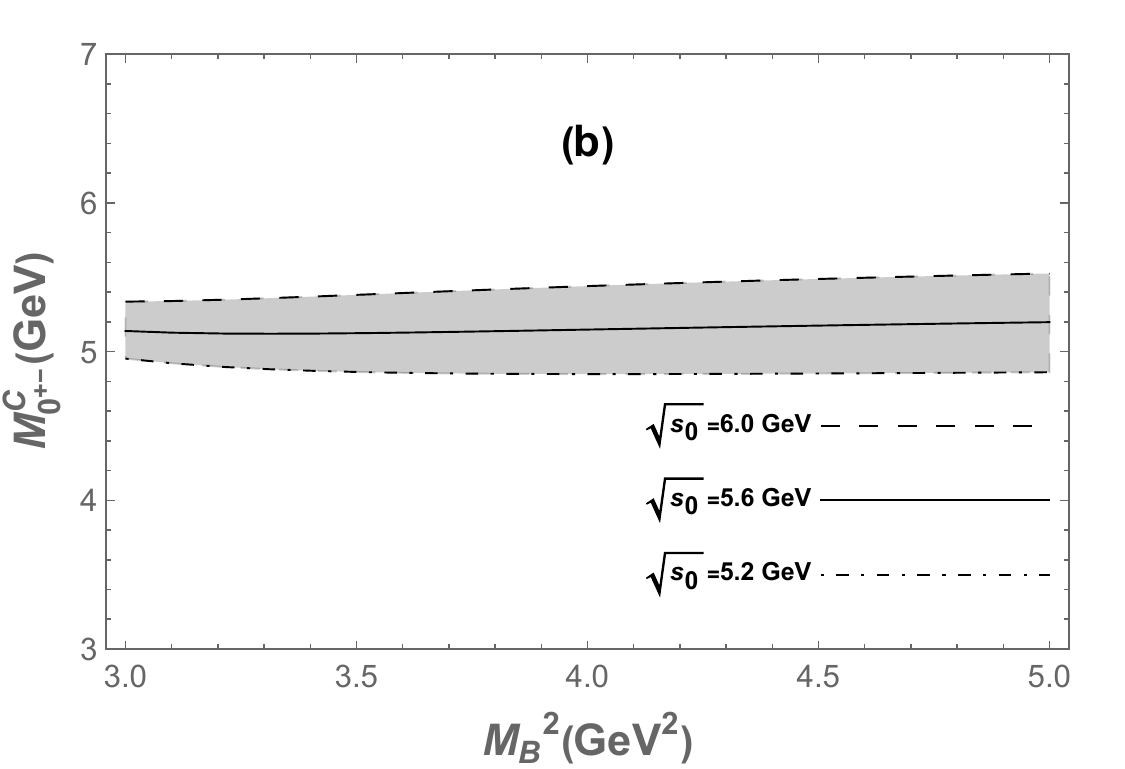}
\caption{The same caption as in Fig \ref{figA0--}, but for the current  in Eq.~(\ref{Jc0+-}).} \label{figC0+-}
\end{figure}

For the $0^{+-}$ hidden-charm exotic baryonium state in Eq.~(\ref{Jc0+-}), the ratios $R^{OPE\;,C}_{0^{+-}}$ and $R^{PC\;,C}_{0^{+-}}$ are presented as functions of Borel parameter $M_B^2$ in Fig. \ref{figC0+-}(a) with different values of $\sqrt{s_0}$, i.e., $5.2$, $5.6$ and $5.9$ GeV. The reliant relations of $M^{C}_{0^{+-}}$ on parameter $M_B^2$ are displayed in Fig. \ref{figC0+-}(b). The optimal Borel window is found in range $3.3 \le M_B^2 \le 4.3\; \text{GeV}^2$, and the mass $M^{C}_{0^{+-}}$ can then be obtained:
\begin{eqnarray}
M^{C}_{0^{+-}} &=& (5.16\pm 0.27)\; \text{GeV}.\label{m6}
\end{eqnarray}

\begin{figure}[h]
\includegraphics[width=6.8cm]{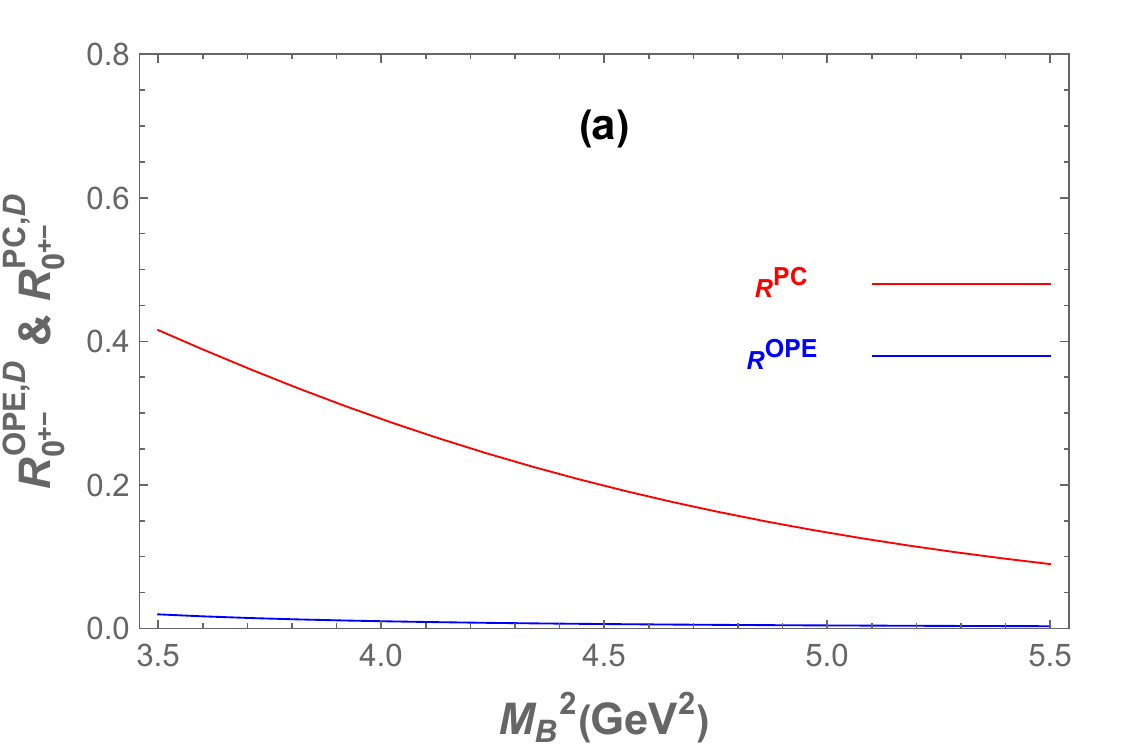}
\includegraphics[width=6.8cm]{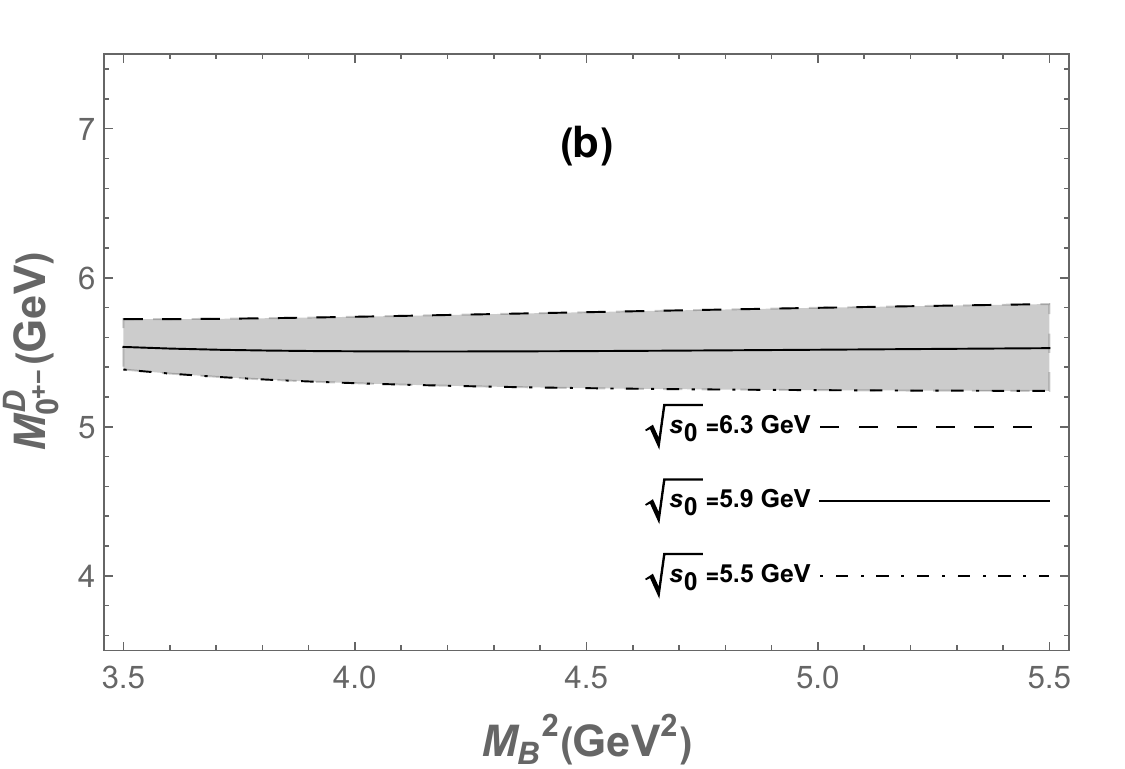}
\caption{The same caption as in Fig \ref{figA0--}, but for the current  in Eq.~(\ref{Jd0+-}).} \label{figD0+-}
\end{figure}

For the $0^{+-}$ hidden-charm exotic baryonium state in Eq.~(\ref{Jd0+-}), the ratios $R^{OPE\;,D}_{0^{+-}}$ and $R^{PC\;,D}_{0^{+-}}$ are presented as functions of Borel parameter $M_B^2$ in Fig. \ref{figD0+-}(a) with different values of $\sqrt{s_0}$, i.e., $5.5$, $5.9$ and $6.3$ GeV. The reliant relations of $M^{D}_{0^{+-}}$ on parameter $M_B^2$ are displayed in Fig. \ref{figD0+-}(b). The optimal Borel window is found in range $3.9 \le M_B^2 \le 4.9\; \text{GeV}^2$, and the mass $M^{D}_{0^{+-}}$ can then be obtained:
\begin{eqnarray}
M^{D}_{0^{+-}} &=& (5.52\pm 0.27)\; \text{GeV}.\label{m7}
\end{eqnarray}

\begin{figure}[h]
\includegraphics[width=6.8cm]{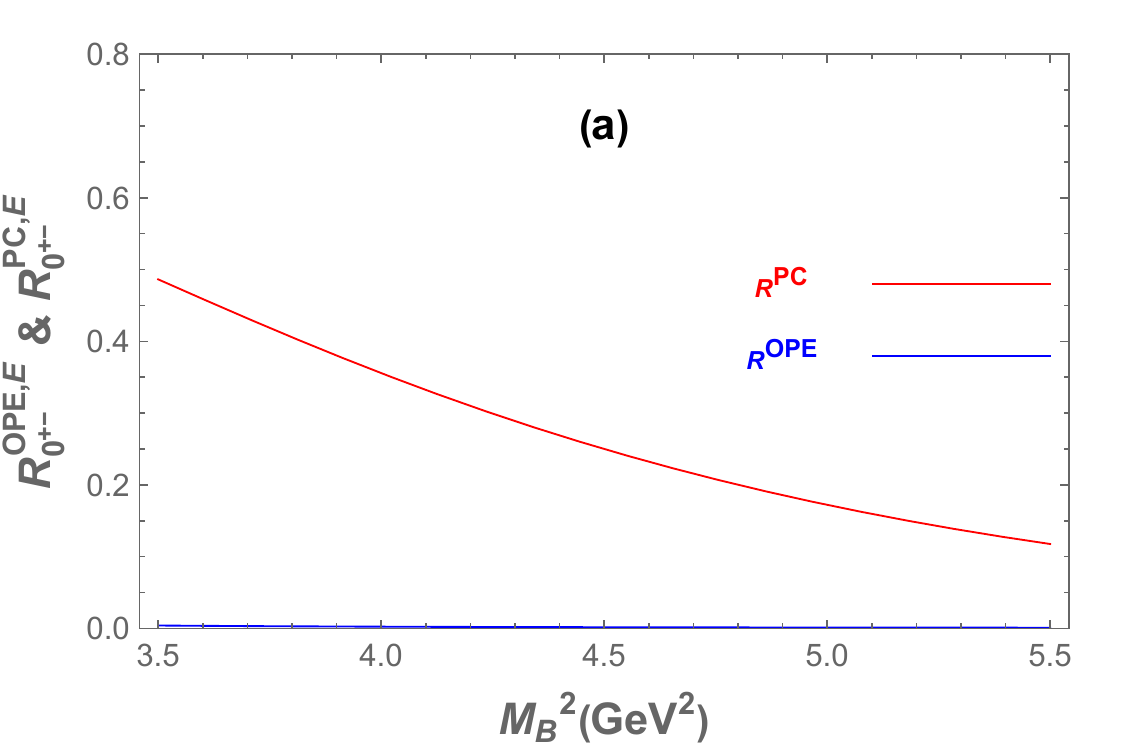}
\includegraphics[width=6.8cm]{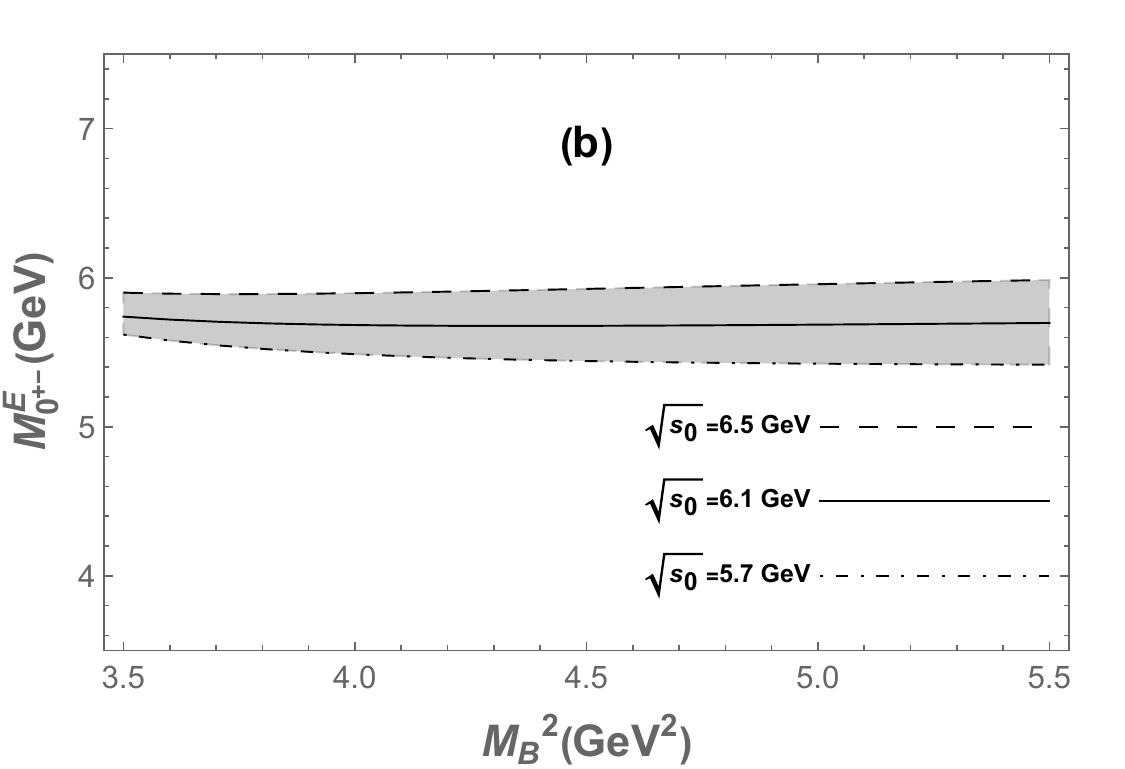}
\caption{The same caption as in Fig \ref{figA0--}, but for the current  in Eq.~(\ref{Je0+-}).} \label{figE0+-}
\end{figure}

For the $0^{+-}$ hidden-charm exotic baryonium state in Eq.~(\ref{Je0+-}), the ratios $R^{OPE\;,E}_{0^{+-}}$ and $R^{PC\;,E}_{0^{+-}}$ are presented as functions of Borel parameter $M_B^2$ in Fig. \ref{figE0+-}(a) with different values of $\sqrt{s_0}$, i.e., $5.7$, $6.1$ and $6.5$ GeV. The reliant relations of $M^{E}_{0^{+-}}$ on parameter $M_B^2$ are displayed in Fig. \ref{figE0+-}(b). The optimal Borel window is found in range $3.8 \le M_B^2 \le 5.2\; \text{GeV}^2$, and the mass $M^{E}_{0^{+-}}$ can then be obtained:
\begin{eqnarray}
M^{E}_{0^{+-}} &=& (5.69\pm 0.27)\; \text{GeV}.\label{m8}
\end{eqnarray}

The $\overline{\text{MS}}$ quark mass are used above. 
The errors in results mainly stem from the uncertainties in quark masses, condensates and threshold parameter $\sqrt{s_0}$. For the currents $j_{0^{--}}^D$, $j_{0^{--}}^E$, $j_{0^{--}}^F$, and $j_{0^{+-}}^F$, no matter what values of $M_B^2$ and $\sqrt{s_0}$ take, no optimal window for stable plateaus exist. That means the currents in Eqs.~(\ref{Jd0--}), (\ref{Je0--}), (\ref{Jf0--}), and (\ref{Jf0+-}) do not support the corresponding hidden-charm baryonium states.
For the convenience of reference, a collection of Borel parameters, continuum thresholds, and predicted masses for hidden-charm exotic baryonium states are tabulated in Table \ref{mass}.

\begin{table}
\begin{center}
\renewcommand\tabcolsep{10pt}
\caption{The continuum thresholds, Borel parameters, and predicted masses of hidden-charm baryonium for $m_c=1.275 \text{GeV}$.}\label{mass}
\begin{tabular}{ccccc}\hline\hline
$J^{PC}$      &Current   & $\sqrt{s_0}\;(\text{GeV})$     &$M_B^2\;(\text{GeV}^2)$ &$M^X\;(\text{GeV})$       \\ \hline
$0^{--}$        &$A$        & $5.6\pm0.4$                             &$3.1-4.1$                      &$5.22\pm0.26$         \\
                     &$B$        & $5.9\pm0.4$                             &$3.2-4.3$                      &$5.52\pm0.25$          \\
                     &$C$        & $5.8\pm0.4$                             &$4.1-5.1$                      &$5.46\pm0.24$ \\\hline
$0^{+-}$       &$A$        & $5.1\pm0.4$                           &$2.8-3.8$                     &$4.76\pm0.28$           \\
                     &$B$        & $5.6\pm0.4$                           &$3.0-4.1$                     &$5.24\pm0.28$           \\
                     &$C$        & $5.6\pm0.4$                          &$3.3-4.3$                      &$5.16\pm0.27$          \\      
                     &$D$        & $5.9\pm0.4$                          &$3.9-4.9$                      &$5.52\pm0.27$          \\  
                     &$E$        & $6.1\pm0.4$                          &$3.8-5.2$                      &$5.69\pm0.27$          \\                                                           
\hline
 \hline
\end{tabular}
\end{center}
\end{table}

\begin{figure}[h]
\includegraphics[width=6.8cm]{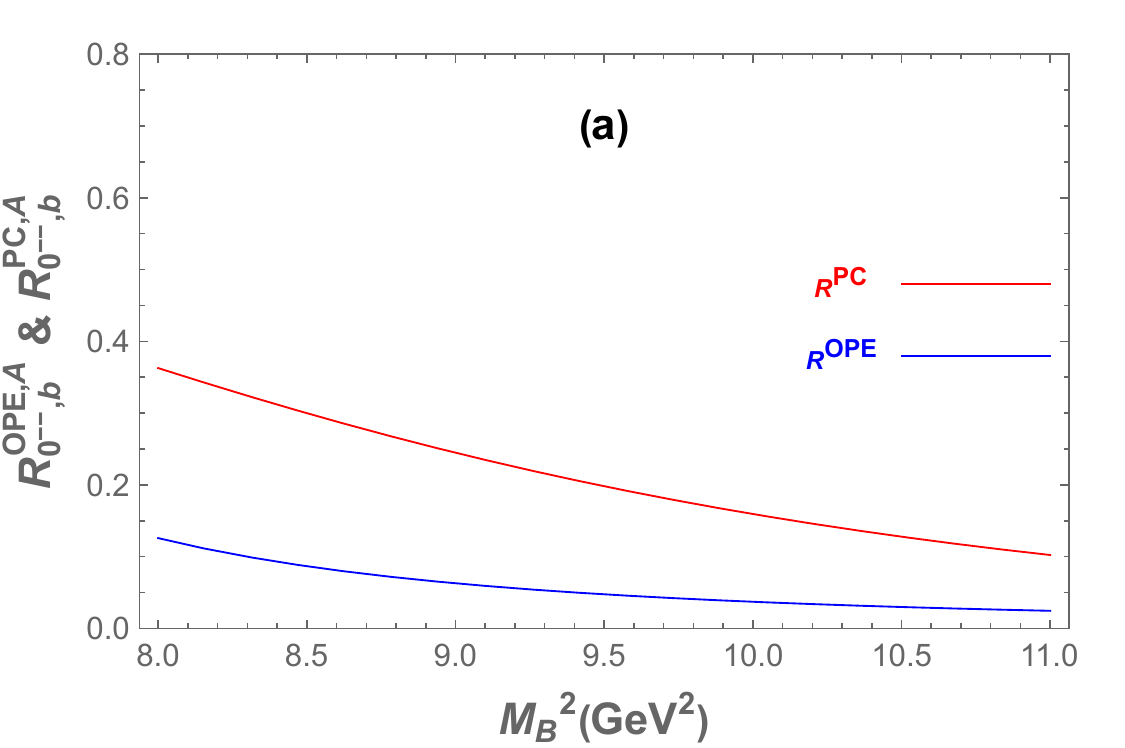}
\includegraphics[width=6.8cm]{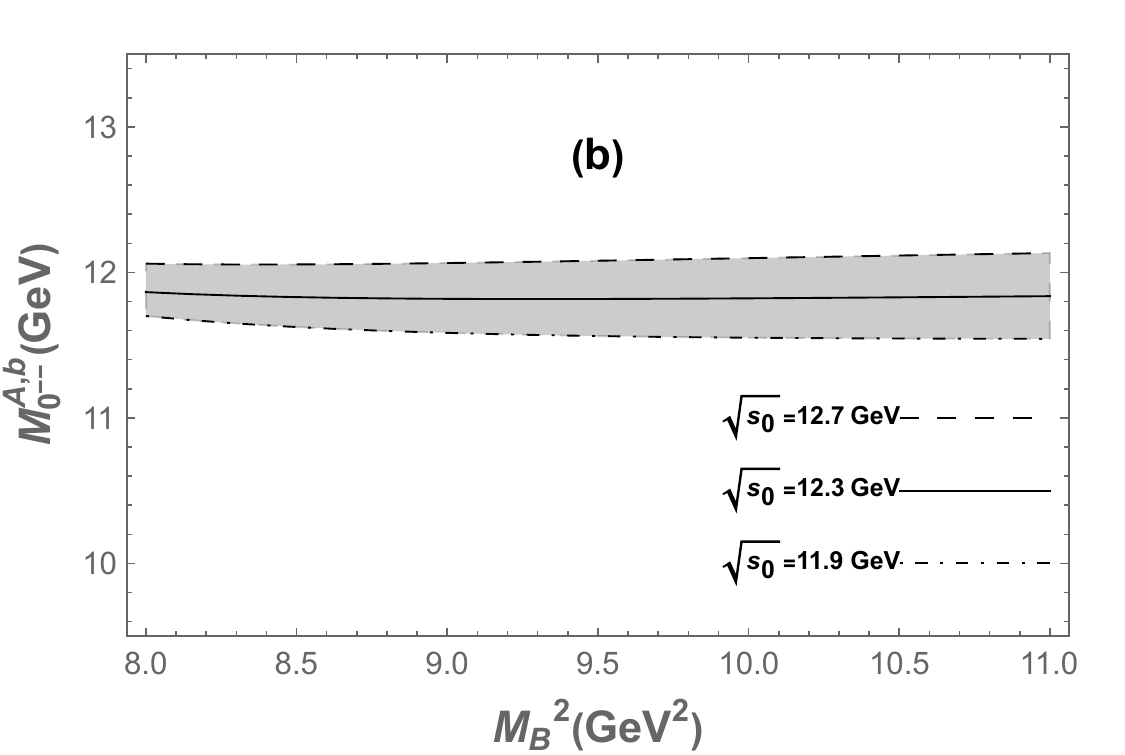}
\caption{The same caption as in Fig \ref{figA0--}, but for the hidden-bottom baryonium state with current in Eq.~(\ref{Ja0--}).} \label{figAb0--}
\end{figure}

 For the $0^{--}$ hidden-bottom exotic baryonium state in Eq.~(\ref{Ja0--}),  the ratios $R^{OPE\;,A}_{0^{--}\;,b}$ and $R^{PC\;,A}_{0^{--}\;,b}$ are presented as functions of Borel parameter $M_B^2$ in Fig. \ref{figAb0--}(a) with different values of $\sqrt{s_0}$, i.e., $11.9$, $12.3$ and $12.7$ GeV. The reliant relations of $M^{A}_{0^{--}\;,b}$ on parameter $M_B^2$ are displayed in Fig. \ref{figAb0--}(b). The optimal Borel window is found in range $8.3 \le M_B^2 \le 10.2\; \text{GeV}^2$, and the mass $M^{A}_{0^{--}\;,b}$ can then be obtained:
\begin{eqnarray}
M^{A}_{0^{--}\;,b} &=& (11.82\pm 0.27)\; \text{GeV}.\label{m9}
\end{eqnarray}

\begin{figure}[h]
\includegraphics[width=6.8cm]{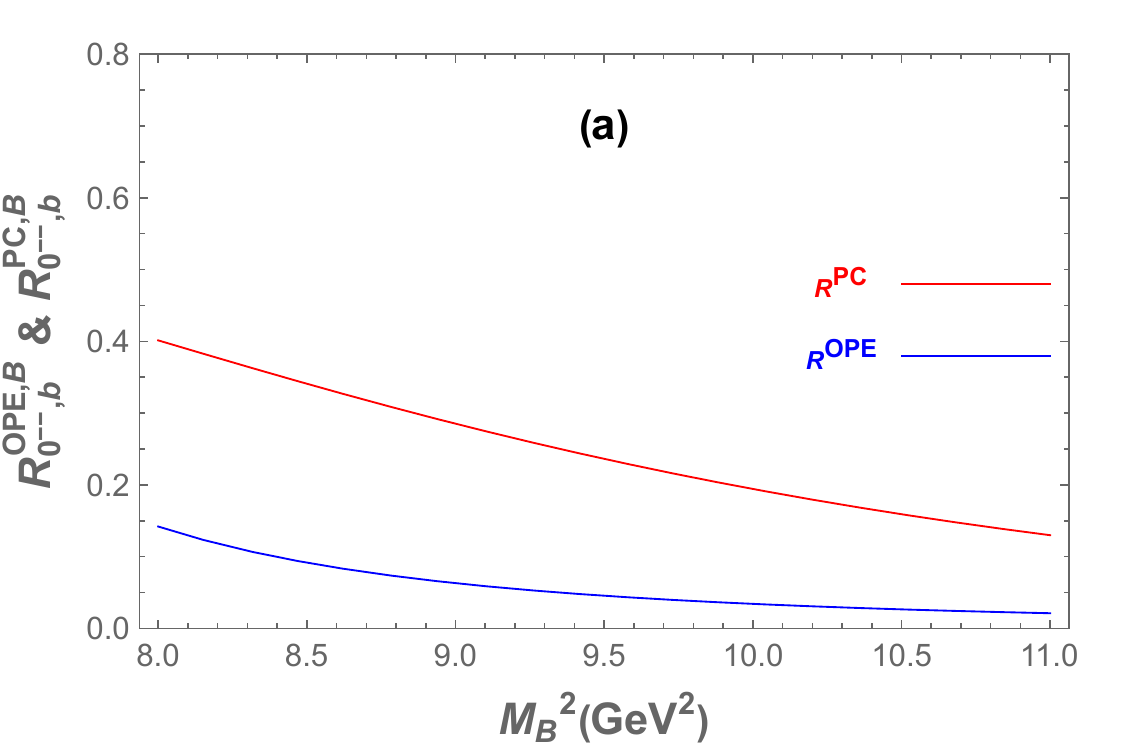}
\includegraphics[width=6.8cm]{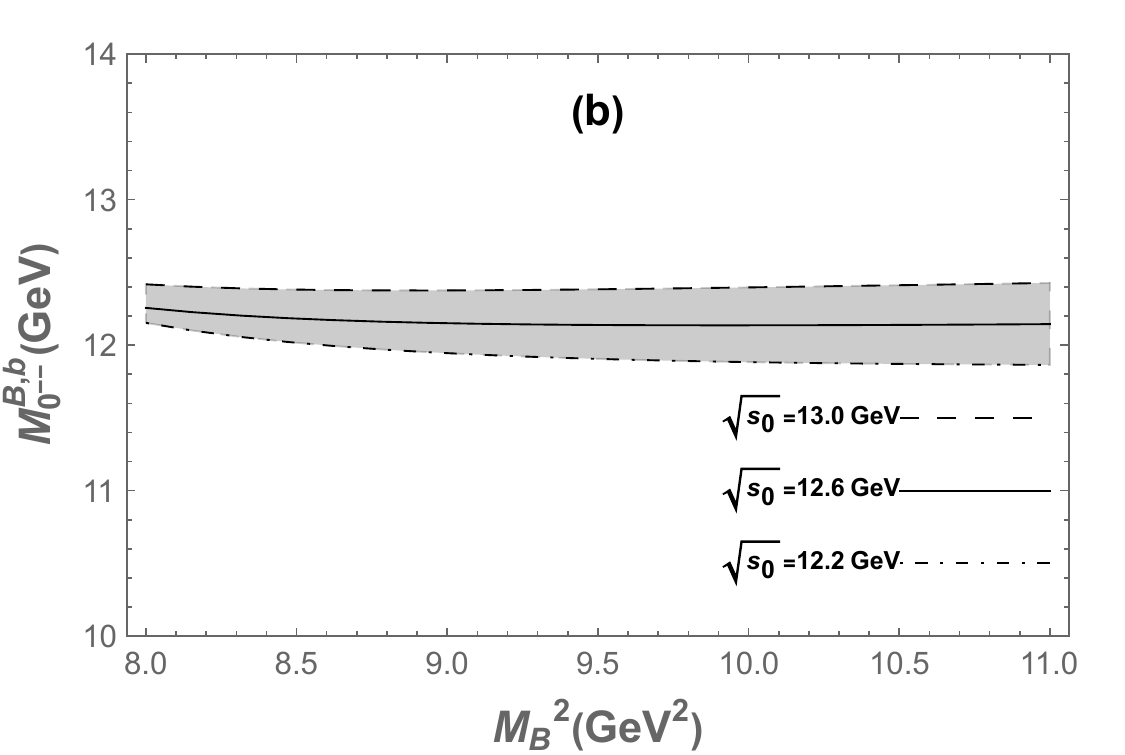}
\caption{The same caption as in Fig \ref{figA0--}, but for the hidden-bottom baryonium state with current in Eq.~(\ref{Jb0--}).} \label{figBb0--}
\end{figure}

 For the $0^{--}$ hidden-bottom exotic baryonium state in Eq.~(\ref{Jb0--}),  the ratios $R^{OPE\;,B}_{0^{--}\;,b}$ and $R^{PC\;,B}_{0^{--}\;,b}$ are presented as functions of Borel parameter $M_B^2$ in Fig. \ref{figBb0--}(a) with different values of $\sqrt{s_0}$, i.e., $12.2$, $12.6$ and $13.0$ GeV. The reliant relations of $M^{B}_{0^{--}\;,b}$ on parameter $M_B^2$ are displayed in Fig. \ref{figBb0--}(b). The optimal Borel window is found in range $8.5 \le M_B^2 \le 10.6\; \text{GeV}^2$, and the mass $M^{B}_{0^{--}\;,b}$ can then be obtained:
\begin{eqnarray}
M^{B}_{0^{--}\;,b} &=& (12.14\pm 0.26)\; \text{GeV}.\label{m10}
\end{eqnarray}

\begin{figure}[h]
\includegraphics[width=6.8cm]{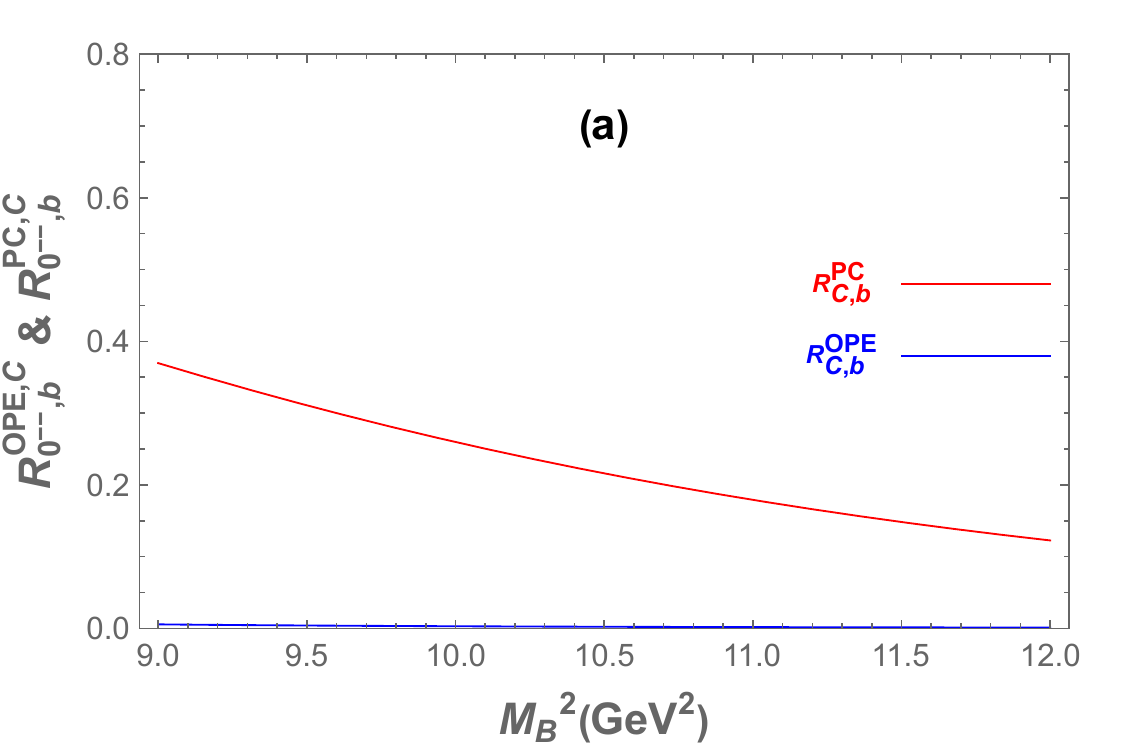}
\includegraphics[width=6.8cm]{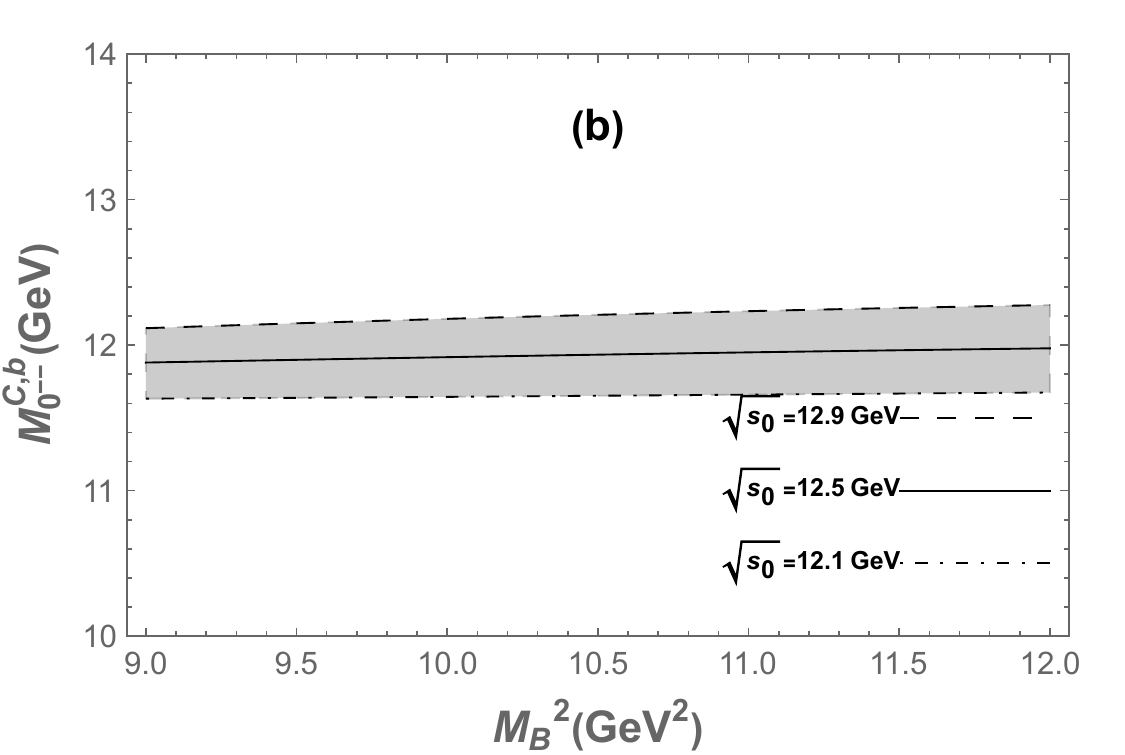}
\caption{The same caption as in Fig \ref{figA0--}, but for the hidden-bottom baryonium state with current in Eq.~(\ref{Jc0--}).} \label{figCb0--}
\end{figure}

 For the $0^{--}$ hidden-bottom exotic baryonium state in Eq.~(\ref{Jc0--}),  the ratios $R^{OPE\;,C}_{0^{--}\;,b}$ and $R^{PC\;,C}_{0^{--}\;,b}$ are presented as functions of Borel parameter $M_B^2$ in Fig. \ref{figCb0--}(a) with different values of $\sqrt{s_0}$, i.e., $12.1$, $12.5$ and $12.9$ GeV. The reliant relations of $M^{C}_{0^{--}\;,b}$ on parameter $M_B^2$ are displayed in Fig. \ref{figCb0--}(b). The optimal Borel window is found in range $9.5 \le M_B^2 \le 12.5\; \text{GeV}^2$, and the mass $M^{C}_{0^{--}\;,b}$ can then be obtained:
\begin{eqnarray}
M^{C}_{0^{--}\;,b} &=& (11.95\pm 0.31)\; \text{GeV}.\label{m11}
\end{eqnarray}

\begin{figure}[h]
\includegraphics[width=6.8cm]{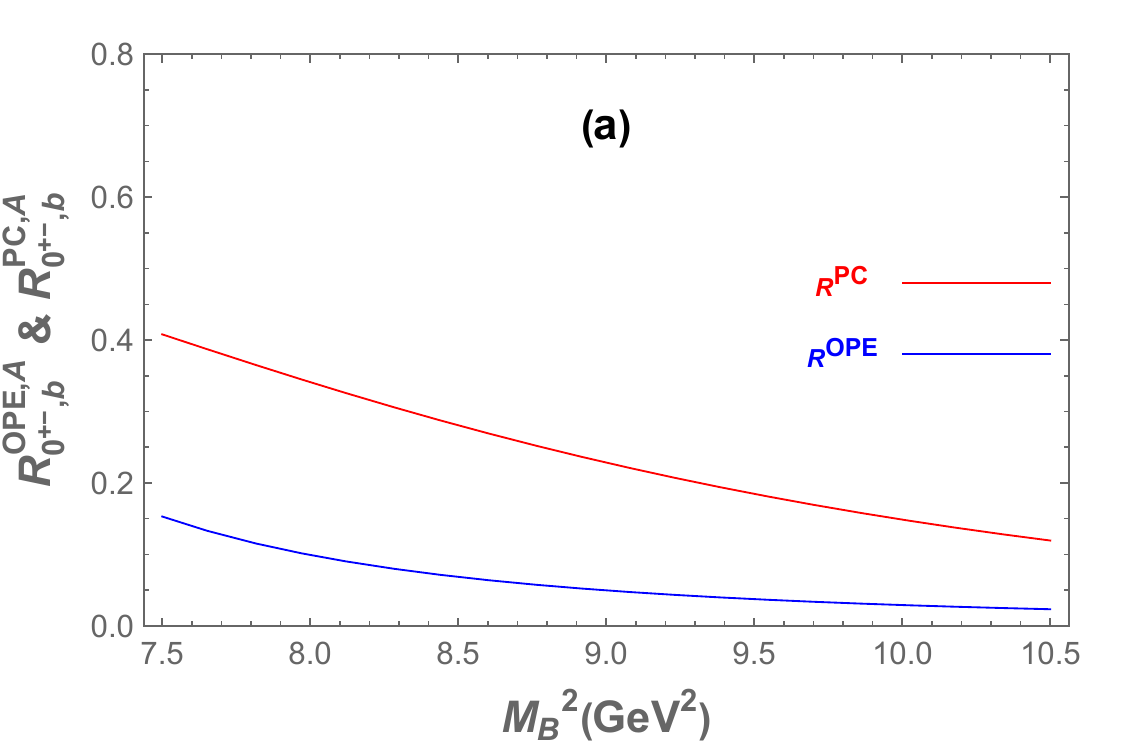}
\includegraphics[width=6.8cm]{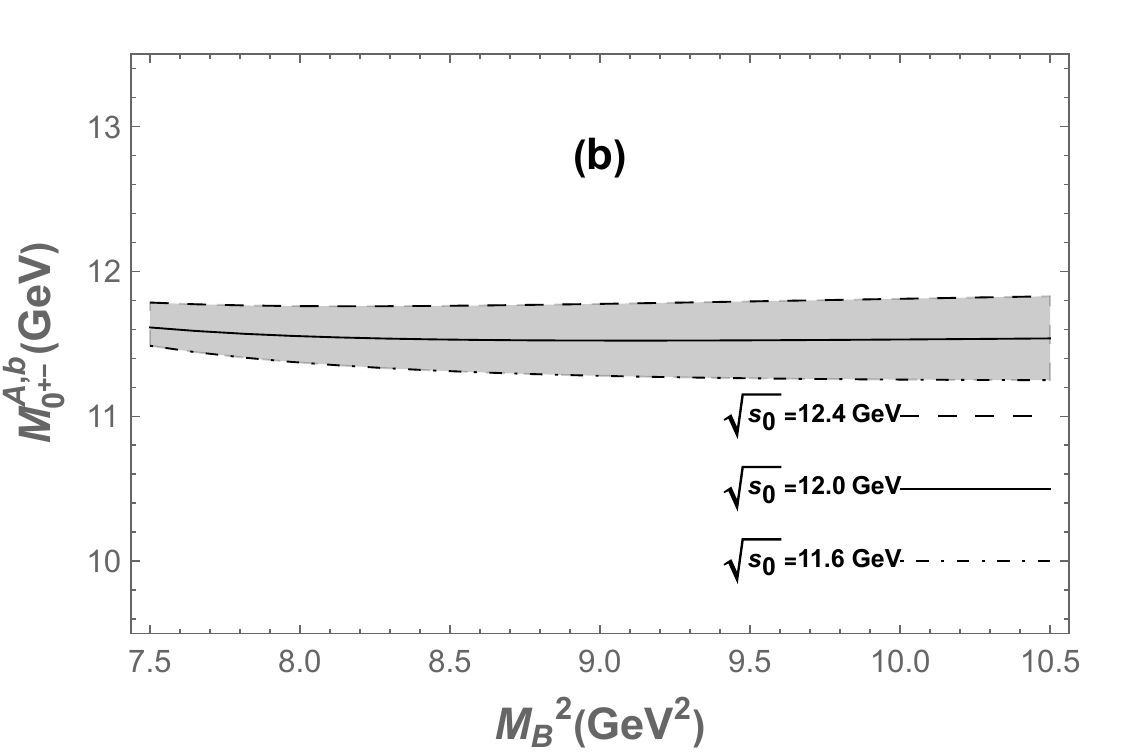}
\caption{The same caption as in Fig \ref{figA0--}, but for the hidden-bottom baryonium state with current in Eq.~(\ref{Ja0+-}).} \label{figAb0+-}
\end{figure}

 For the $0^{+-}$ hidden-bottom exotic baryonium state in Eq.~(\ref{Ja0+-}),  the ratios $R^{OPE\;,A}_{0^{+-}\;,b}$ and $R^{PC\;,A}_{0^{+-}\;,b}$ are presented as functions of Borel parameter $M_B^2$ in Fig. \ref{figAb0+-}(a) with different values of $\sqrt{s_0}$, i.e., $11.6$, $12.0$ and $12.4$ GeV. The reliant relations of $M^{A}_{0^{+-}\;,b}$ on parameter $M_B^2$ are displayed in Fig. \ref{figAb0--}(b). The optimal Borel window is found in range $8.0 \le M_B^2 \le 10.0\; \text{GeV}^2$, and the mass $M^{A}_{0^{+-}\;,b}$ can then be obtained:
\begin{eqnarray}
M^{A}_{0^{+-}\;,b} &=& (11.53\pm 0.27)\; \text{GeV}.\label{m12}
\end{eqnarray}

\begin{figure}[h]
\includegraphics[width=6.8cm]{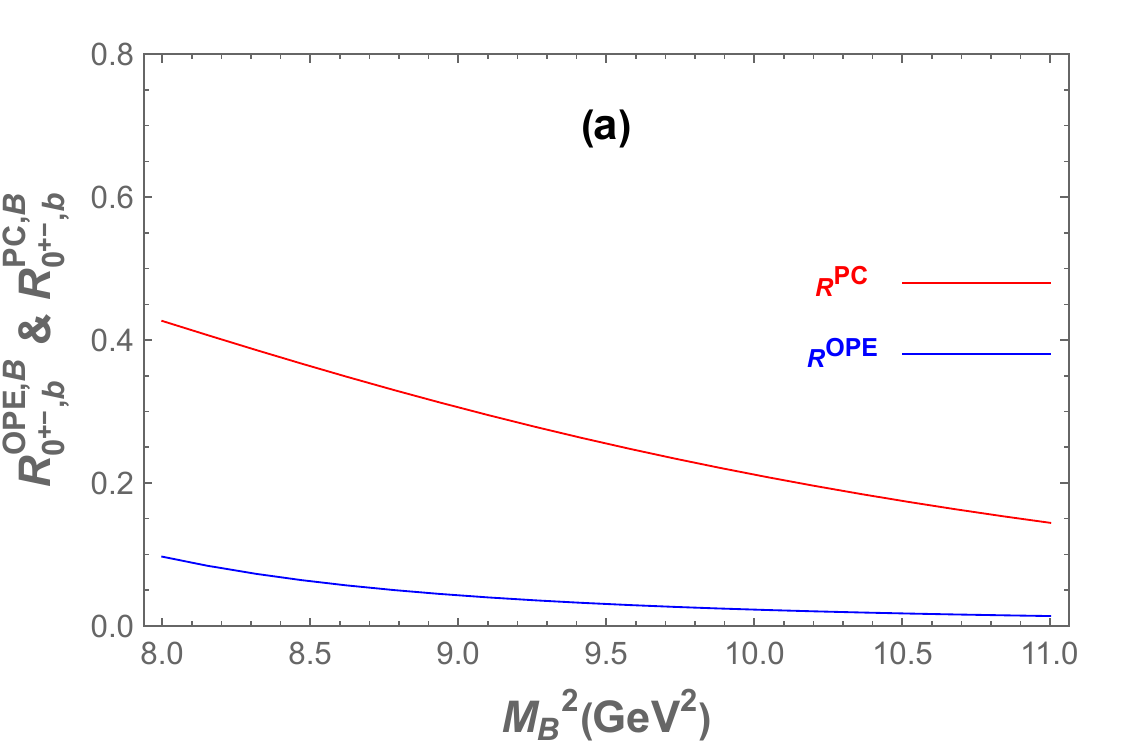}
\includegraphics[width=6.8cm]{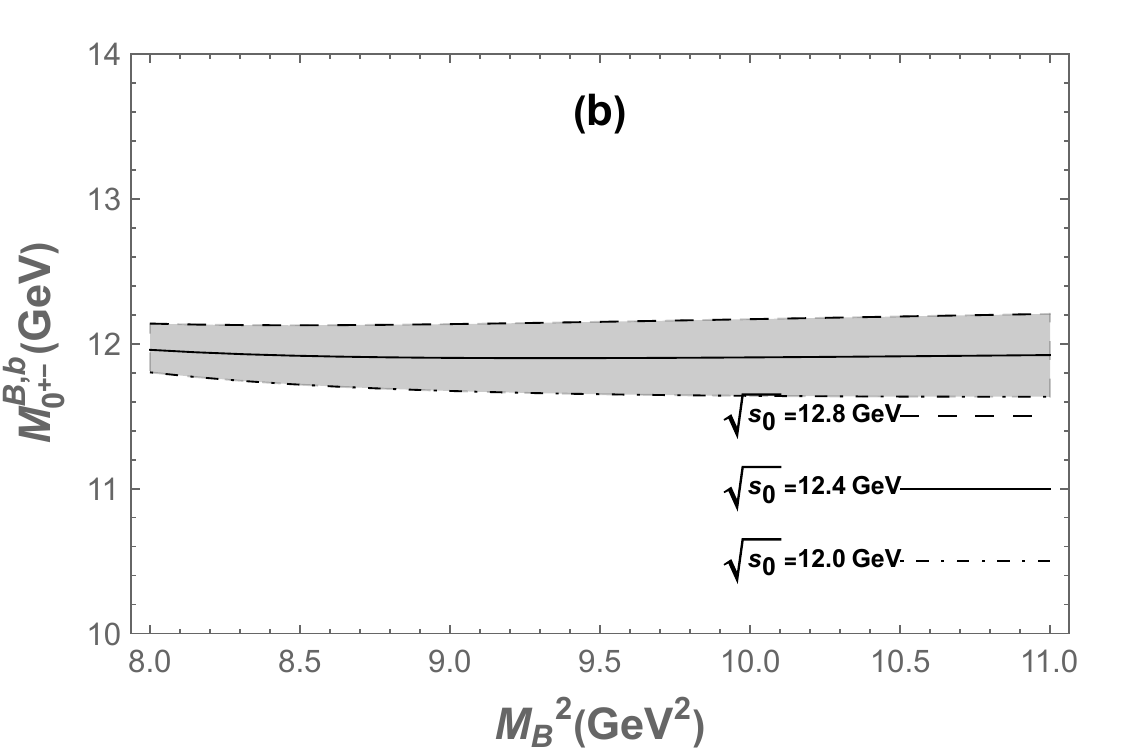}
\caption{The same caption as in Fig \ref{figA0--}, but for the hidden-bottom baryonium state with current in Eq.~(\ref{Jb0+-}).} \label{figBb0+-}
\end{figure}

 For the $0^{+-}$ hidden-bottom exotic baryonium state in Eq.~(\ref{Jb0+-}),  the ratios $R^{OPE\;,B}_{0^{+-}\;,b}$ and $R^{PC\;,B}_{0^{+-}\;,b}$ are presented as functions of Borel parameter $M_B^2$ in Fig. \ref{figBb0+-}(a) with different values of $\sqrt{s_0}$, i.e., $12.0$, $12.4$ and $12.8$ GeV. The reliant relations of $M^{B}_{0^{+-}\;,b}$ on parameter $M_B^2$ are displayed in Fig. \ref{figBb0--}(b). The optimal Borel window is found in range $8.5 \le M_B^2 \le 10.9\; \text{GeV}^2$, and the mass $M^{B}_{0^{+-}\;,b}$ can then be obtained:
\begin{eqnarray}
M^{B}_{0^{+-}\;,b} &=& (11.92\pm 0.28)\; \text{GeV}.\label{m13}
\end{eqnarray}

\begin{figure}[h]
\includegraphics[width=6.8cm]{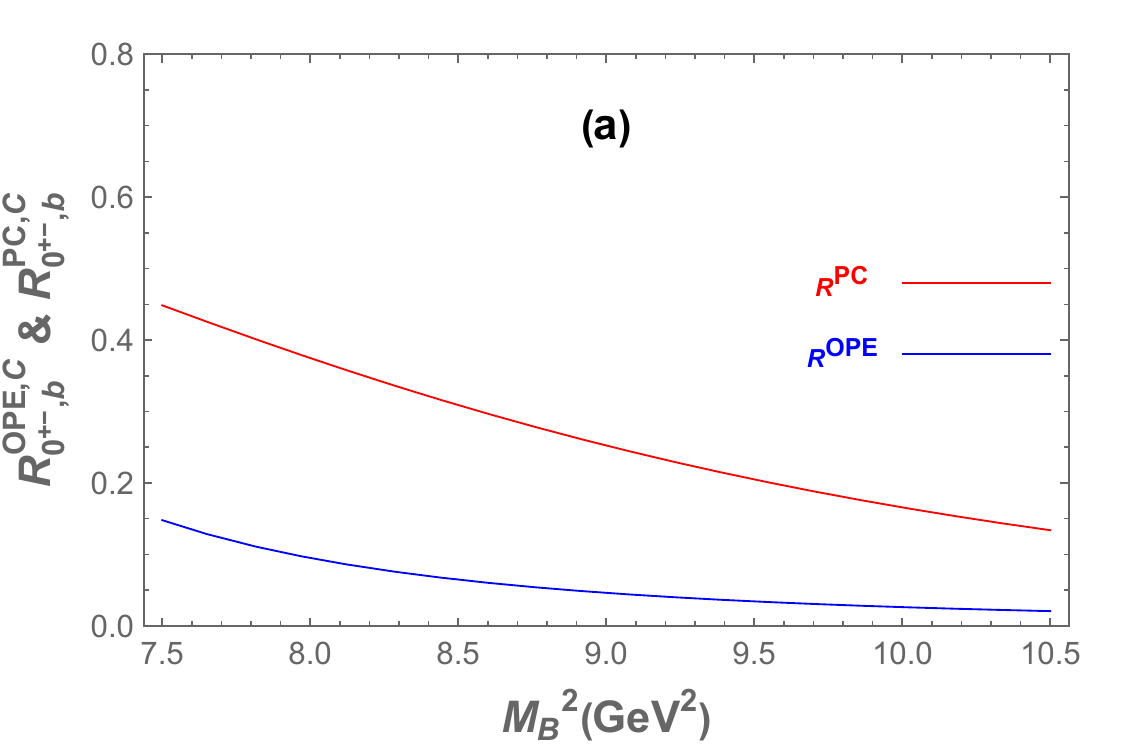}
\includegraphics[width=6.8cm]{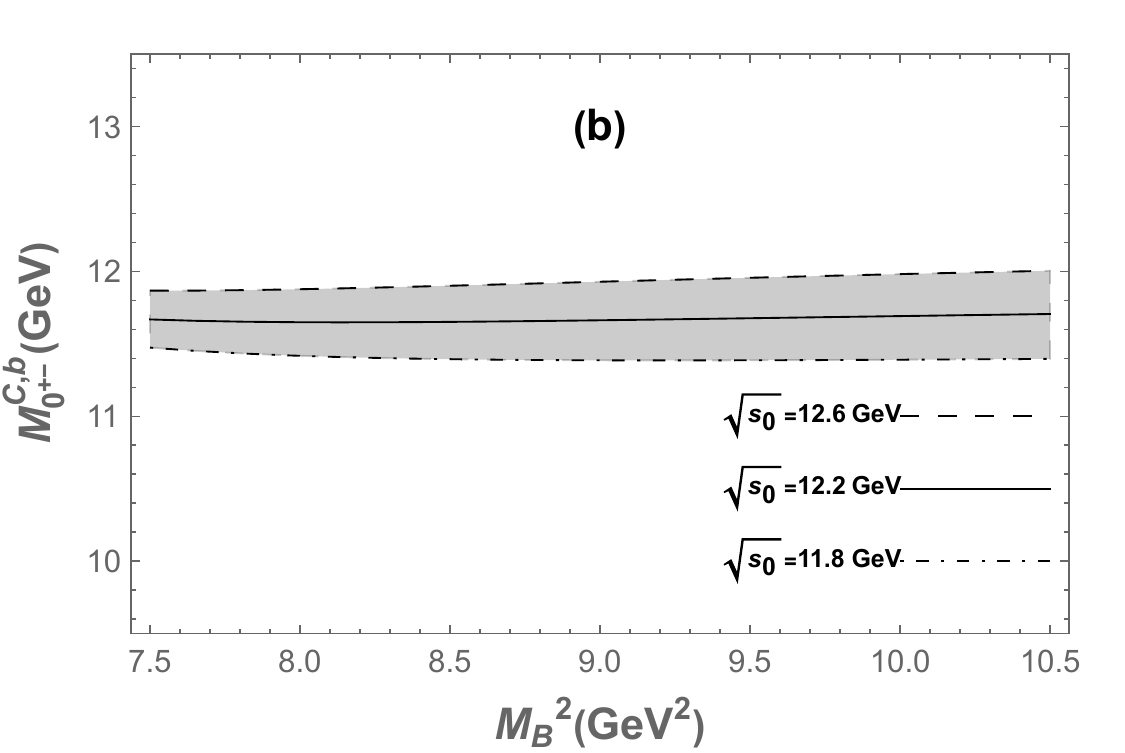}
\caption{The same caption as in Fig \ref{figA0--}, but for the hidden-bottom baryonium state with current in Eq.~(\ref{Jc0+-}).} \label{figCb0+-}
\end{figure}

 For the $0^{+-}$ hidden-bottom exotic baryonium state in Eq.~(\ref{Jc0+-}),  the ratios $R^{OPE\;,C}_{0^{+-}\;,b}$ and $R^{PC\;,C}_{0^{+-}\;,b}$ are presented as functions of Borel parameter $M_B^2$ in Fig. \ref{figCb0+-}(a) with different values of $\sqrt{s_0}$, i.e., $11.8$, $12.2$ and $12.6$ GeV. The reliant relations of $M^{C}_{0^{+-}\;,b}$ on parameter $M_B^2$ are displayed in Fig. \ref{figCb0--}(b). The optimal Borel window is found in range $7.9 \le M_B^2 \le 10.3\; \text{GeV}^2$, and the mass $M^{C}_{0^{+-}\;,b}$ can then be obtained:
\begin{eqnarray}
M^{C}_{0^{+-}\;,b} &=& (11.69\pm 0.29)\; \text{GeV}.\label{m14}
\end{eqnarray}

\begin{figure}[h]
\includegraphics[width=6.8cm]{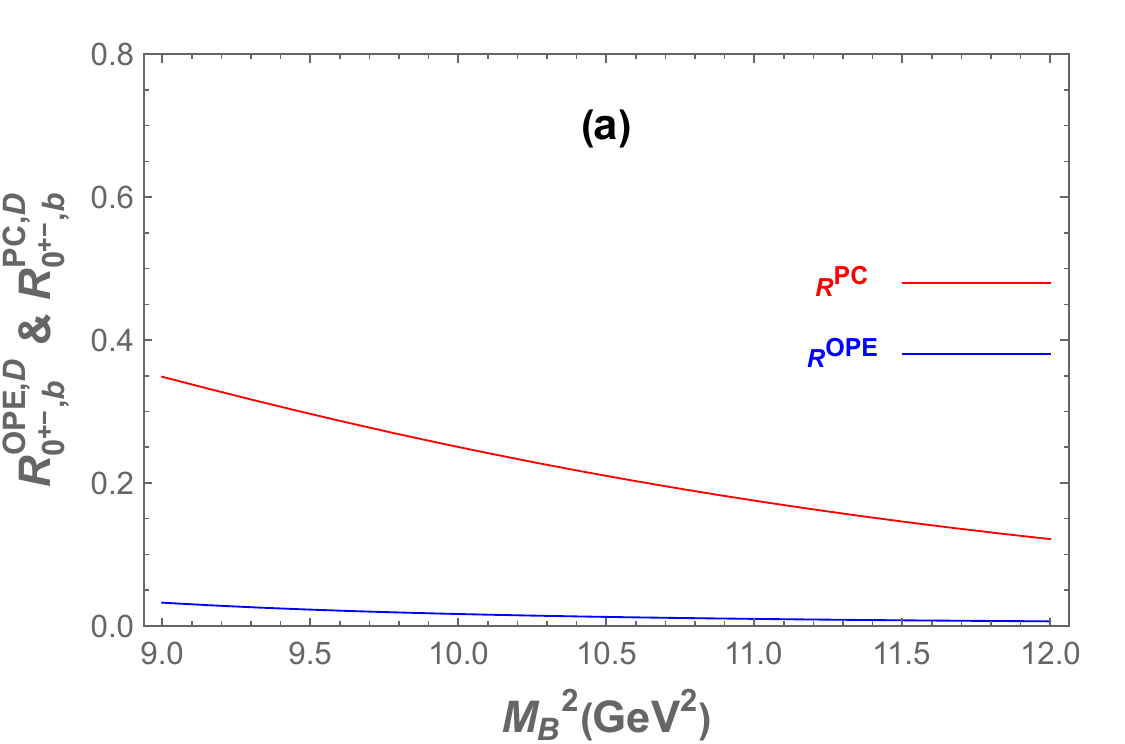}
\includegraphics[width=6.8cm]{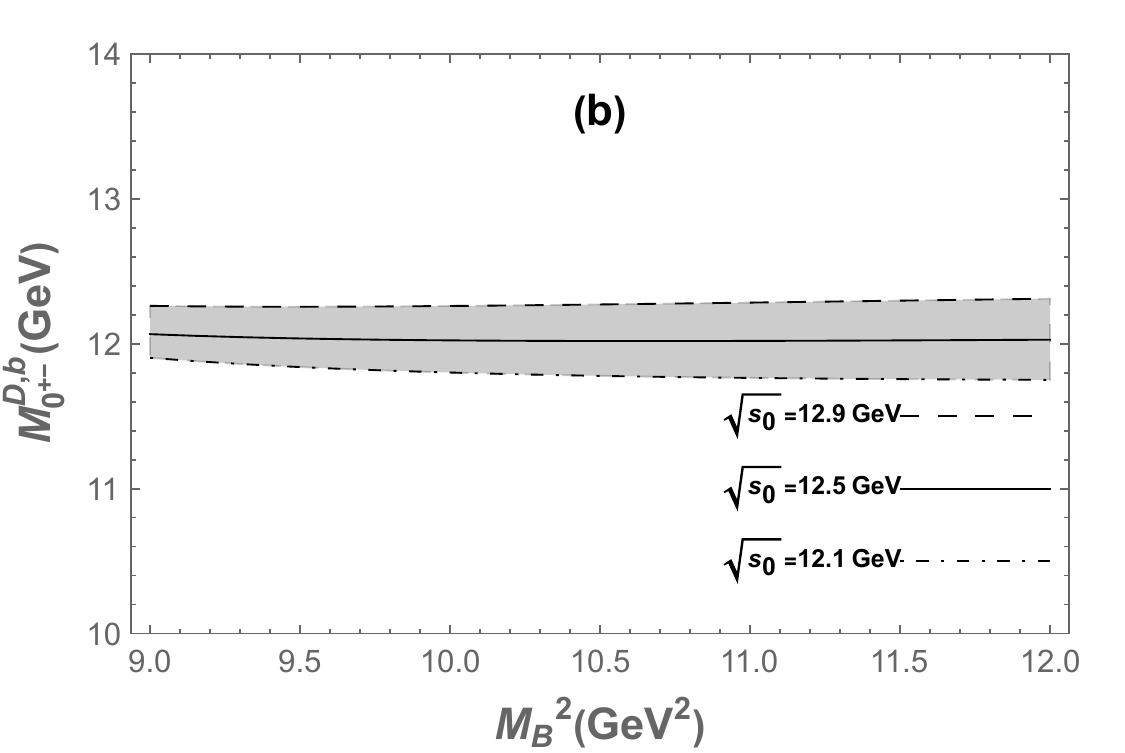}
\caption{The same caption as in Fig \ref{figA0--}, but for the hidden-bottom baryonium state with current in Eq.~(\ref{Jd0+-}).} \label{figDb0+-}
\end{figure}

 For the $0^{+-}$ hidden-bottom exotic baryonium state in Eq.~(\ref{Jd0+-}),  the ratios $R^{OPE\;,D}_{0^{+-}\;,b}$ and $R^{PC\;,D}_{0^{+-}\;,b}$ are presented as functions of Borel parameter $M_B^2$ in Fig. \ref{figDb0+-}(a) with different values of $\sqrt{s_0}$, i.e., $12.1$, $12.5$ and $12.9$ GeV. The reliant relations of $M^{D}_{0^{+-}\;,b}$ on parameter $M_B^2$ are displayed in Fig. \ref{figBb0--}(b). The optimal Borel window is found in range $9.4 \le M_B^2 \le 11.4\; \text{GeV}^2$, and the mass $M^{D}_{0^{+-}\;,b}$ can then be obtained:
\begin{eqnarray}
M^{D}_{0^{+-}\;,b} &=& (12.02\pm 0.26)\; \text{GeV}.\label{m15}
\end{eqnarray}

\begin{figure}[h]
\includegraphics[width=6.8cm]{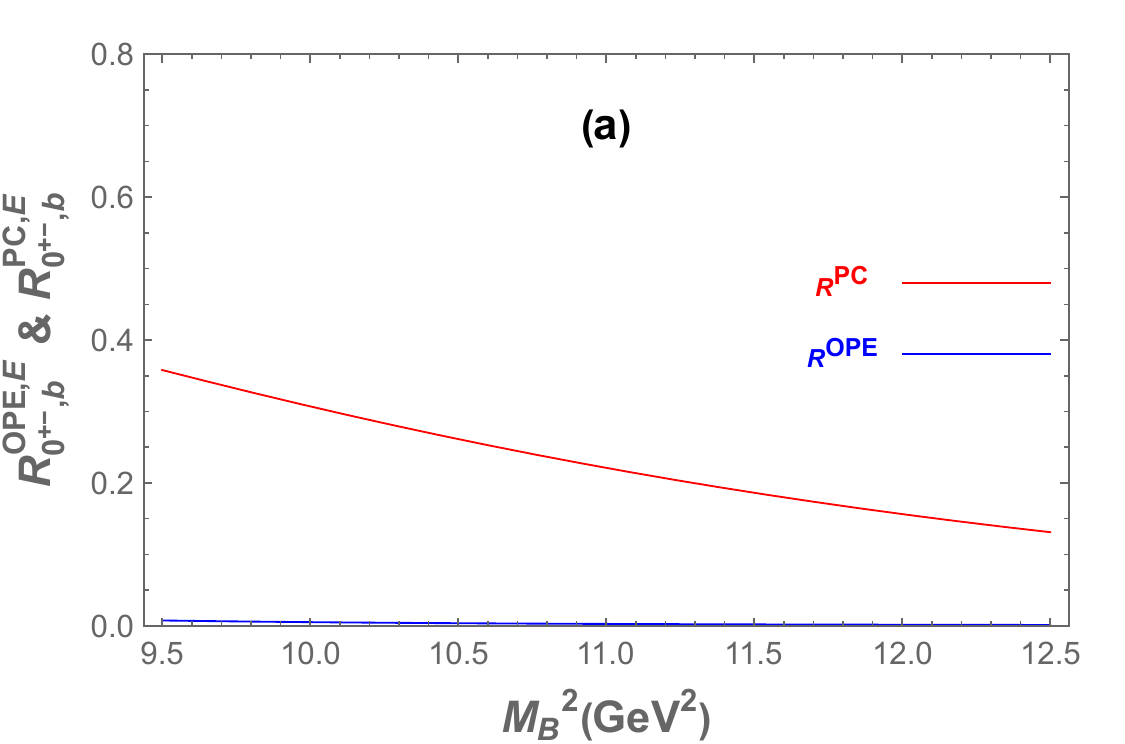}
\includegraphics[width=6.8cm]{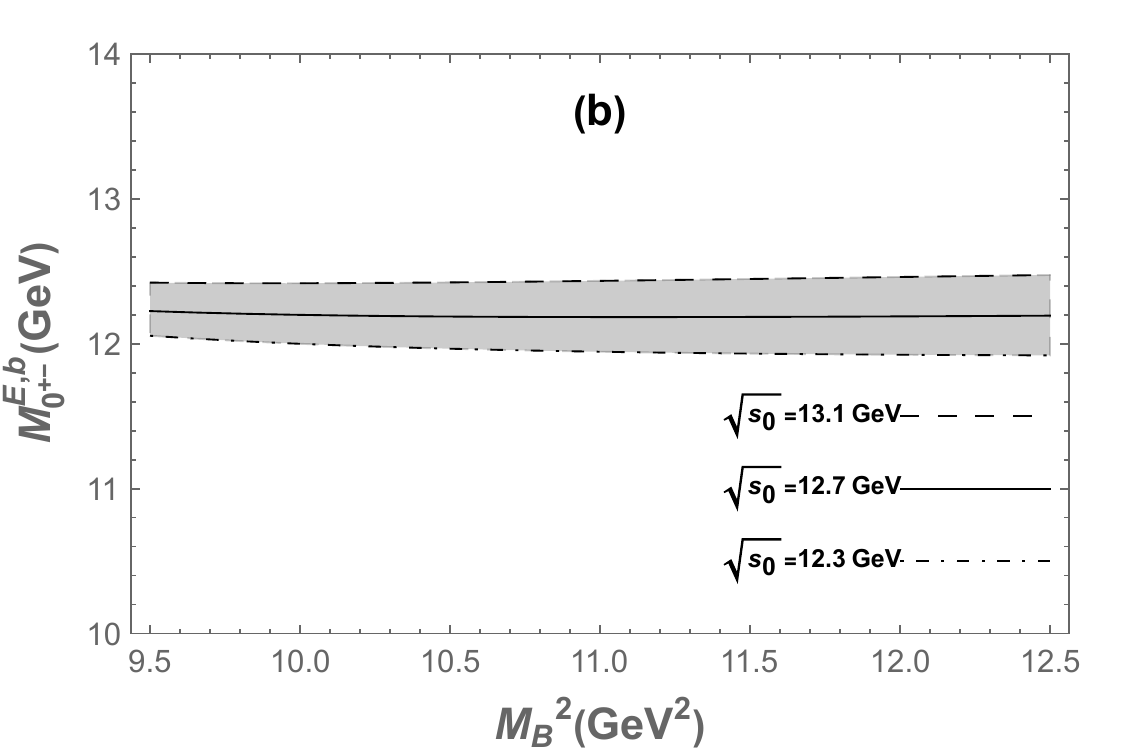}
\caption{The same caption as in Fig \ref{figA0--}, but for the hidden-bottom baryonium state with current in Eq.~(\ref{Je0+-}).} \label{figEb0+-}
\end{figure}

 For the $0^{+-}$ hidden-bottom exotic baryonium state in Eq.~(\ref{Je0+-}),  the ratios $R^{OPE\;,E}_{0^{+-}\;,b}$ and $R^{PC\;,E}_{0^{+-}\;,b}$ are presented as functions of Borel parameter $M_B^2$ in Fig. \ref{figEb0+-}(a) with different values of $\sqrt{s_0}$, i.e., $12.3$, $12.7$ and $13.1$ GeV. The reliant relations of $M^{E}_{0^{+-}\;,b}$ on parameter $M_B^2$ are displayed in Fig. \ref{figEb0--}(b). The optimal Borel window is found in range $10.0 \le M_B^2 \le 12.1\; \text{GeV}^2$, and the mass $M^{E}_{0^{+-}\;,b}$ can then be obtained:
\begin{eqnarray}
M^{E}_{0^{+-}\;,b} &=& (12.19\pm 0.27)\; \text{GeV}.\label{m16}
\end{eqnarray}

The $\overline{\text{MS}}$ quark mass are used above.
The errors in results mainly stem from the uncertainties in quark masses, condensates and threshold parameter $\sqrt{s_0}$. For the currents $j_{0^{--}}^D$, $j_{0^{--}}^E$, $j_{0^{--}}^F$, and $j_{0^{+-}}^F$, no matter what values of $M_B^2$ and $\sqrt{s_0}$ take, no optimal window for stable plateaus exist. That means the currents in Eqs.~(\ref{Jd0--}), (\ref{Je0--}), (\ref{Jf0--}), and (\ref{Jf0+-}) do not support the corresponding hidden-bottom baryonium states.
For the convenience of reference, a collection of Borel parameters, continuum thresholds, and predicted masses for hidden-charm exotic baryonium states are tabulated in Table \ref{mass_b}.

\begin{table}
\begin{center}
\renewcommand\tabcolsep{10pt}
\caption{The continuum thresholds, Borel parameters, and predicted masses of hidden-bottom baryonium for $m_b=4.18 \text{GeV}$.}\label{mass_b}
\begin{tabular}{ccccc}\hline\hline
$J^{PC}$        &Current   & $\sqrt{s_0}\;(\text{GeV})$     &$M_B^2\;(\text{GeV}^2)$ &$M^X\;(\text{GeV})$       \\ \hline
$0^{--}$          &$A$        & $12.3\pm0.4$                           &$8.3-10.2$                    &$11.82\pm0.27$           \\
                       &$B$        & $12.6\pm0.4$                           &$8.5-10.6$                     &$12.14\pm0.26$           \\
                       &$C$        & $12.5\pm0.4$                          &$9.5-12.5$                      &$11.95\pm0.31$          \\\hline
$0^{+-}$          &$A$        & $12.0\pm0.4$                           &$8.0-10.0$                     &$11.53\pm0.27$           \\
                       &$B$        & $12.4\pm0.4$                           &$8.5-10.9$                     &$11.92\pm0.28$           \\
                       &$C$        & $12.2\pm0.4$                          &$7.9-10.3$                      &$11.69\pm0.29$          \\      
                       &$D$        & $12.5\pm0.4$                          &$9.4-11.4$                      &$12.02\pm0.26$          \\ 
                       &$E$        & $12.7\pm0.4$                          &$10.0-12.1$                     &$12.19\pm0.27$          \\                                                             
\hline
 \hline
\end{tabular}
\end{center}
\end{table}

We also evaluate the exotic hidden-charm and hidden-bottom baryonium states with the pole quark masses. By using the obtained analytical results but with pole quark masses, the corresponding masses are readily obtained and tabulated in Table~\ref{mass_bigc} and Table~\ref{mass_b_bigc}.

\begin{table}
\begin{center}
\renewcommand\tabcolsep{10pt}
\caption{The continuum thresholds, Borel parameters, and predicted masses of hidden-charm baryonium for $m_c\approx 1.5 \text{GeV}$.}\label{mass_bigc}
\begin{tabular}{ccccc}\hline\hline
$J^{PC}$      &Current   & $\sqrt{s_0}\;(\text{GeV})$     &$M_B^2\;(\text{GeV}^2)$ &$M^X\;(\text{GeV})$       \\ \hline
$0^{--}$        &$A$        & $5.7\pm0.4$                             &$3.2-4.2$                      &$5.39\pm0.26$         \\
                     &$B$        & $6.1\pm0.4$                             &$3.2-4.3$                      &$5.77\pm0.25$          \\
                     &$C$        & $5.9\pm0.4$                             &$4.1-5.1$                      &$5.63\pm0.24$ \\\hline
$0^{+-}$       &$A$        & $5.2\pm0.4$                           &$2.8-3.8$                     &$4.95\pm0.26$           \\
                     &$B$        & $5.7\pm0.4$                           &$3.0-4.1$                     &$5.45\pm0.28$           \\
                     &$C$        & $5.7\pm0.4$                          &$3.3-4.3$                      &$5.37\pm0.27$          \\      
                     &$D$        & $5.9\pm0.4$                          &$3.9-4.9$                      &$5.69\pm0.27$          \\  
                     &$E$        & $6.2\pm0.4$                          &$3.8-5.2$                      &$5.86\pm0.28$          \\                                                           
\hline
 \hline
\end{tabular}
\end{center}
\end{table}

\begin{table}
\begin{center}
\renewcommand\tabcolsep{10pt}
\caption{The continuum thresholds, Borel parameters, and predicted masses of hidden-bottom baryonium for $m_b\approx 4.7 \text{GeV}$.}\label{mass_b_bigb}
\begin{tabular}{ccccc}\hline\hline
$J^{PC}$        &Current   & $\sqrt{s_0}\;(\text{GeV})$     &$M_B^2\;(\text{GeV}^2)$ &$M^X\;(\text{GeV})$       \\ \hline
$0^{--}$          &$A$        & $12.4\pm0.4$                           &$8.5-10.3$                    &$12.15\pm0.23$           \\
                       &$B$        & $12.7\pm0.4$                           &$8.5-10.6$                     &$12.49\pm0.22$           \\
                       &$C$        & $12.6\pm0.4$                          &$9.5-12.5$                      &$12.29\pm0.27$          \\\hline
$0^{+-}$          &$A$        & $12.0\pm0.4$                           &$8.0-10.0$                     &$11.87\pm0.27$           \\
                       &$B$        & $12.4\pm0.4$                           &$8.5-10.9$                     &$12.27\pm0.28$           \\
                       &$C$        & $12.3\pm0.4$                          &$7.9-10.3$                      &$12.04\pm0.29$          \\      
                       &$D$        & $12.6\pm0.4$                          &$9.4-11.4$                      &$12.41\pm0.26$          \\ 
                       &$E$        & $12.8\pm0.4$                          &$10.0-12.1$                     &$12.56\pm0.28$          \\                                                             
\hline
 \hline
\end{tabular}
\end{center}
\end{table}

\section{Decay analyses}\label{decay}

To finally ascertain these exotic hidden-heavy baryonium states, the straightforward procedure is to reconstruct them from their decay products, though the detailed characters stills ask for more investigation. The typical decay modes of these exotic baryonium are given in Table \ref{decay}, and these processes are expected to be measurable in the running LHC experiments.

\begin{table}
\begin{center}
\renewcommand\tabcolsep{10pt}
\caption{typical decay modes of the exotic hidden-heavy baryonium.}\label{decay}
\begin{tabular}{ccccc}\hline\hline
$J^{PC}$             & $c$-sector                                                                & $b$-sector      \\ \hline
$0^{--}$               & $\Lambda_c^+ \Lambda_c(2595)^-$                     &$\Lambda_b \overline{\Lambda}_b(5912)$   \\
                             & $\Lambda_c^- \Lambda_c(2595)^+$                     &$\overline{\Lambda}_b \Lambda_b(5912)$\\
                             & $\Lambda_c(2625)^+ \Lambda_c(2860)^-$           &$\Sigma^\ast_b \overline{\Lambda}_b(5920)$   \\
                             & $\Lambda_c(2625)^- \Lambda_c(2860)^+$           &$\overline{\Sigma}^\ast_b \Lambda_b(5920)$\\
                            & $\Lambda_c^+ \Lambda_c(2860)^-\gamma$          &$\Sigma^\ast_b \overline{\Lambda}_b\gamma$                               \\
                            &$\Lambda_c^- \Lambda_c(2860)^+\gamma$          &$\overline{\Sigma}_b \Lambda_b\gamma$ \\\hline
$0^{+-}$             & $\Lambda_c^+ \Lambda_c(2595)^-$                     &$\Lambda_b \overline{\Lambda}_b(5912)$   \\
                            & $\Lambda_c^- \Lambda_c(2595)^+$                     &$\overline{\Lambda}_b \Lambda_b(5912)$\\
                            & $\Lambda_c(2625)^+ \Lambda_c(2860)^-$           &$\Sigma^\ast_b \overline{\Lambda}_b(5920)$   \\
                            & $\Lambda_c(2625)^- \Lambda_c(2860)^+$           &$\overline{\Sigma}^\ast_b \Lambda_b(5920)$\\
                            & $\Lambda_c(2595)^+ \Sigma_c(2520)^-\gamma$  & $\Lambda_b \overline{\Lambda}_b(5920)\gamma$                              \\
                            &$\Lambda_c(2595)^- \Sigma_c(2520)^+\gamma$    &$\overline{\Lambda}_b \Lambda_b(5920)\gamma$ \\
                           & $\Lambda_c^+ \Lambda_c(2625)^-\gamma$           &$\Sigma^\ast_b \overline{\Lambda}_b\gamma$                            \\ 
                           &$\Lambda_c^- \Lambda_c(2625)^+\gamma$            &$\overline{\Sigma}^\ast_b \Lambda_b\gamma$ \\
                            & $\Lambda_c^+ \Lambda_c(2860)^-\gamma$           &                           \\ 
                           &$\Lambda_c^- \Lambda_c(2860)^+\gamma$            &\\                        
\hline
 \hline
\end{tabular}
\end{center}
\end{table}

\section{Summary}

In summary, we have investigated the exotic hidden-heavy baryonium states with $J^{PC}=0^{--}$ and $0^{+-}$ in the framework of QCD sum rules. Our numerical results are tabulated in Table \ref{mass} and \ref{mass_b} for charm and bottom sector, respectively. Results indicate that there might exist 3 possible $0^{--}$ hidden-charm baryonium states with masses $(5.22\pm0.26)$, $(5.52\pm0.25)$, and $(5.46\pm0.24)$ GeV, and 5 possible $0^{+-}$ hidden-charm baryonium states with masses $(4.76\pm0.28)$, $(5.24\pm0.28)$, $(5.16\pm0.27)$, $(5.52\pm0.27)$, and $(5.69\pm0.27)$ GeV, respectively. The corresponding hidden-bottom partners are found lying in the range of $11.68-12.28$ GeV and $11.38-12.33$ GeV, respectively. Moreover, the possible exotic baryonium decay modes are analyzed, which might serve as a guide for experimental exploration.

It should noted that, as the discussion in Ref. \cite{Albuquerque:2016znh}, lowest order of $\overline{\text{MS}}$ quark masses is ill-defined such that one can equally use the running or the pole masses. One way to solve this problem is to consider the next-to leading and next-to-next-to leading order QCD corrections in the analysis. While, in the work, we just naively perform our calculation by taking consideration on both $\overline{\text{MS}}$ quark mass and the pole masses. The numerical results by using the pole masses are tabulated in Table~\ref{mass_bigc} and Table~\ref{mass_b_bigb}, which is bigger than the results in tabulated in Table~\ref{mass} and Table~\ref{mass_b}.

It also should be noted that we list six $0^{--}$ and six $0^{+-}$ different interpolating operators exotic baryonium states, while they are very different states. From the structures of the interpolating operators, it's not hard to see that, for the $0^{--}$ states the Eq. (\ref{Ja0--}) is coupled to $\Lambda_c\bar{\Lambda}_c(2595)-c.c.$ molecular state, Eq. (\ref{Jb0--}) is coupled to $\Lambda_c(2625)\bar{\Lambda}_c(2860)-c.c.$ molecular state, Eq. (\ref{Jc0--}) is coupled to $\Lambda_c\bar{\Lambda}_c(2860)+c.c.$ molecular state, Eq. (\ref{Jd0--}) is coupled to $\Lambda_c(2625)\bar{\Lambda}_c(2595)+c.c.$ molecular state, the Eq. (\ref{Je0--}) is coupled to $\Lambda_c\bar{\Lambda}_c(2625)-c.c.$ molecular state, and the Eq. (\ref{Jf0--}) is coupled to $\Lambda_c(2595)\bar{\Lambda}_c(2860)-c.c.$ molecular state. On the other hand, for the $0^{+-}$ states the Eq. (\ref{Ja0+-}) is coupled to $\Lambda_c\bar{\Lambda}_c(2595)-c.c.$ molecular state, Eq. (\ref{Jb0+-}) is coupled to $\Lambda_c(2625)\bar{\Lambda}_c(2860)-c.c.$ molecular state, Eq. (\ref{Jc0+-}) is coupled to $\Lambda_c(2595)\bar{\Lambda}_c(2860)+c.c.$ molecular state, Eq. (\ref{Jd0+-}) is coupled to $\Lambda_c\bar{\Lambda}_c(2625)+c.c.$ molecular state, Eq. (\ref{Je0+-}) is coupled to $\Lambda_c(2595)\bar{\Lambda}_c(2625)-c.c.$ molecular state, and Eq. (\ref{Jf0+-}) is coupled to $\Lambda_c\bar{\Lambda}_c(2860)-c.c.$ molecular state. Thus the experiments can discriminate those states.

%%%%%%%%%%%%%%%%%%%%%%%%%%%%%%%%%%%%%%%%%%%%%%%%%%%%%%%%%%%%%%%%%%%%%%
\vspace{.5cm} {\bf Acknowledgments} \vspace{.5cm}

This work was supported in part by the National Natural Science Foundation of China (NSFC) under the Grants 12247113, Project funded by China Postdoctoral Science Foundation under the Grants 2022M723117, Specific Fund of Fundamental Scientific Research Operating Expenses for Undergraduate Universities in Liaoning Province (2024) and Ph.~D. Research Start-up Fund of Liaoning Normal University (no. 2024BSL026).

%%%%%%%%%%%%%%%%%%%%%%%%%%%%%%%%%%%%%%%%%%%%%%%%%%%%%%%%%%%%%%%%%%%%%%%

\begin{widetext}
\appendix

\section{The spectral densities for $0^{--}$ and $0^{+-}$ baryonium states}\label{ana_exp}
\subsection{The spectral densities for $0^{--}$ baryonium state in Eqs.~(\ref{Ja0--}) and (\ref{Jb0--})  }
\begin{eqnarray}
\rho^{pert}(s)&=&\int_{\alpha_{min}}^{\alpha_{max}} d \alpha \int_{\beta_{min}}^{1-\alpha} d \beta \bigg{\{}\frac{{\cal F}_{\alpha \beta}^{6}(\alpha+\beta-1)^{4}(-5{\cal F}_{\alpha \beta}+7m_Q^2(\alpha+\beta-1)}{3\times7\times5^2\times2^{19}\pi^{10} \alpha^{6} \beta^{6}}\bigg{\}}\; ,\\
\rho^{\langle G^2 \rangle}(s)&=&\frac{\langle G^2 \rangle}{\pi^{10}}\int_{\alpha_{min}}^{\alpha_{max}} d \alpha \int_{\beta_{min}}^{1-\alpha} d \beta \bigg{\{}{\cal N}_i \frac{{\cal F}_{\alpha \beta}^{4} (\alpha+\beta-1)^{2} (5 m_Q^{2}(\alpha+\beta-1)-3{\cal F}_{\alpha \beta})}{3\times5\times2^{19} \alpha^{4} \beta^{4}}\nonumber\\
&+&\frac{{\cal F}_{\alpha \beta}^{3}m_Q^{2}(\alpha+\beta-1)^{4}}{5\times3^2\times2^{21} \alpha^{6} \beta^{6}}\left(4 m_Q^{2}\left(\alpha^{4}+\alpha^{3}(\beta-1)+\alpha \beta^{3}+(\beta-1) \beta^{3}\right)\right. \nonumber\\
&-&{\cal F}_{\alpha \beta}\left(2 \alpha^{3}-3 \alpha^{2}(\beta-1)-3 \alpha \beta^{2}+\beta^{2}(3+2 \beta)) \right)\bigg{\}}\; ,\\
\rho^{\langle G^3 \rangle}(s) &=&\frac{-\langle G^3 \rangle}{5\times3^2\times2^{23}\pi^{10}}\int_{\alpha_{min}}^{\alpha_{max}} \frac{d \alpha}{ \alpha ^6} \int_{\beta_{min}}^{1-\alpha} \frac{d \beta}{ \beta ^6} {\cal F}_{\alpha \beta}^2  (\alpha +\beta -1)^4  \nonumber\\
&\times&\bigg{\{}5 {\cal F}_{\alpha \beta}^2(\alpha^3+\beta^3)+8 {\cal F}_{\alpha \beta} m_Q^2 (2 \alpha^4 -3\alpha^3(\beta-1)-3\alpha \beta^3 \nonumber\\
&+&3\beta^3+2\beta^4)-24  m_Q^4 (\alpha +\beta -1)(\alpha^4+\beta^4)\bigg{\}}\; ,\\
\rho^{\langle \bar{q} q\rangle^4}&=&\int_{\alpha_{min}}^{\alpha_{max}} d \alpha \frac{{\cal H}_\alpha}{48 \pi^{2}}\langle \bar{q} q\rangle^4\;,
\end{eqnarray}
where $M_B$ is the Borel parameter introduced by the Borel
transformation, $Q = c$ or $b$, and the factor ${\cal N}_i$ has the following definition : ${\cal N}_A=1$ and ${\cal N}_B=0$. Here, we also have the following definitions:
\begin{eqnarray}
{\cal F}_{\alpha \beta} &=& (\alpha + \beta) m_Q^2 - \alpha \beta s \; , {\cal H}_\alpha  = m_Q^2 - \alpha (1 - \alpha) s \; , \\
\alpha_{min} &=& \left(1 - \sqrt{1 - 4 m_Q^2/s} \right) / 2, \; , \alpha_{max} = \left(1 + \sqrt{1 - 4 m_Q^2 / s} \right) / 2  \; , \\
\beta_{min} &=& \alpha m_Q^2 /(s \alpha - m_Q^2).
\end{eqnarray}

\subsection{The spectral densities for $0^{--}$ baryonium state in Eqs.~(\ref{Jc0--}) and (\ref{Jd0--})  }

\begin{eqnarray}
\rho^{pert} (s) &=&  \int^{\alpha_{max}}_{\alpha_{min}} d \alpha \int^{1 - \alpha}_{\beta_{min}} d \beta  \frac{{\cal F}^6_{\alpha \beta} (\alpha + \beta - 1)^4 \Big(7  m_Q^2 (\alpha + \beta - 1)+ 10{\cal F}_{\alpha \beta} \Big) }{3\times 7 \times 5^2 \times 2^{18}\pi^{10} \alpha^6 \beta^6}\;,\\
\rho^{\langle G^2 \rangle}(s) &=& \frac{\langle g_s^2 G^2\rangle}{\pi^{10}} \int^{\alpha_{max}}_{\alpha_{min}} d \alpha \int^{1 - \alpha}_{\beta_{min}} d \beta \bigg{\{}- \frac{{\cal F}_{\alpha \beta}^4 (\alpha + \beta -1)^2}{3\times5\times2^{21}\alpha^5\beta^5}\Big(5m_Q^2(\alpha+\beta-1)^2\nonumber\\
&\times&(\alpha+\beta)+4{\cal F}_{\alpha \beta} \big(\alpha^2+\beta(\beta-1)-\alpha(4\beta+1)\big)  \Big)
+\frac{{\cal F}_{\alpha \beta}^3 m_Q^2 (\alpha + \beta -1)^4 }{3^2\times5\times2^{20}\alpha^6\beta^6} \nonumber\\
&\times&\bigg ( 4 m_Q^2 \Big(\alpha^4 +\alpha^3(\beta-1) +\alpha\beta^3+\beta^3(\beta-1)\Big)+{\cal F}_{\alpha \beta}\Big(13 \alpha^3 +3 \alpha^2 (\beta-1) \nonumber\\
&+&3\alpha \beta^2 +\beta^2(13\beta-3) \Big) \bigg) \bigg{\}}\;,\\
\rho^{\langle \bar{q} q \rangle^2}(s)&=&{\cal N}_i \frac{\langle \bar{q} q \rangle^2}{2^{10}\pi^6} \int^{\alpha_{max}}_{\alpha_{min}} d \alpha \int^{1 - \alpha}_{\beta_{min}} d \beta \frac{2 m_Q^2 {\cal F}_{\alpha \beta}^3 (\alpha+\beta-1)^2+{\cal F}_{\alpha \beta}^4 (\alpha+\beta - 1)}{\alpha^3\beta^3}\;,\\
\rho^{\langle G^3 \rangle}(s) &=&\frac{\langle G^3 \rangle}{5\times3^2\times2^{21}\pi^{10}}\int_{\alpha_{min}}^{\alpha_{max}} d \alpha \int_{\beta_{min}}^{1-\alpha} d \beta \frac{ {\cal F}_{\alpha \beta}^2  (\alpha +\beta -1)^4}{\alpha^6\beta^6}  \nonumber\\
&\times& \bigg( 5(\alpha^3+\beta^3) {\cal F}_{\alpha \beta}^2+4 {\cal F}_{\alpha \beta} m_Q^2 \Big( 13\alpha^4+3\alpha^3(\beta-1)+3\alpha\beta^3\nonumber\\
&+&\beta^3(13\beta-3) \Big)+12  m_Q^4 \Big(\alpha^5+\alpha^4(\beta-1)+\alpha\beta^4+\beta^4(\beta-1) \Big)\bigg)\; ,\\
\rho^{\langle \bar{q} q \rangle \langle \bar{q} G q \rangle}(s)&=&\frac{-{\cal N}_i\langle \bar{q} q \rangle \langle \bar{q} G q \rangle}{2^9\pi^6}\int_{\alpha_{min}}^{\alpha_{max}} d \alpha \int_{\beta_{min}}^{1-\alpha} d \beta \frac{{\cal F}_{\alpha \beta}^{3}+3 m_Q^{2}{\cal F}_{\alpha \beta}^2(\alpha+\beta-1)}{ \alpha^{2} \beta^{2}}\; ,\\
\rho^{\langle \bar{q} G q \rangle^2}&=&\frac{{\cal N}_i\langle \bar{q} G q \rangle^2}{2^{11}\pi^6}\int_{\alpha_{min}}^{\alpha_{max}} d \alpha \bigg{\{} \frac{3{\cal H}_\alpha^{2} }{2 (1-\alpha) \alpha}+\int_{\beta_{min}}^{1-\alpha} d \beta \frac{3{\cal F}_{\alpha \beta} m_Q^{2} }{ \alpha \beta}\bigg{\}} \; ,\\
\rho^{\langle \bar{q} q\rangle^4}&=&\int_{\alpha_{min}}^{\alpha_{max}} d \alpha \frac{{\cal H}_\alpha-m_Q^2}{24 \pi^{2}}\langle \bar{q} q\rangle^4\;,
\end{eqnarray}
where ${\cal N}_C=1$ and ${\cal N}_D=-1$.

\subsection{The spectral densities for $0^{--}$ baryonium state in Eqs.~(\ref{Je0--}) and (\ref{Jf0--})}

\begin{eqnarray}
\rho^{pert} (s) &=&  \int^{\alpha_{max}}_{\alpha_{min}} d \alpha \int^{1 - \alpha}_{\beta_{min}} d \beta  \frac{{\cal F}^6_{\alpha \beta} (\alpha + \beta - 1)^4 \Big( 10{\cal F}_{\alpha \beta}-7  m_Q^2 (\alpha + \beta - 1) \Big) }{3\times 7 \times 5^2 \times 2^{18}\pi^{10} \alpha^6 \beta^6}\;,\\
\rho^{\langle G^2 \rangle}(s) &=& \frac{\langle g_s^2 G^2\rangle}{\pi^{10}} \int^{\alpha_{max}}_{\alpha_{min}} d \alpha \int^{1 - \alpha}_{\beta_{min}} d \beta \bigg{\{} \frac{{\cal F}_{\alpha \beta}^4 (\alpha + \beta -1)^2}{3\times5\times2^{21}\alpha^5\beta^5}\Big(5m_Q^2(\alpha+\beta-1)^2\nonumber\\
&\times&(\alpha+\beta)-4{\cal F}_{\alpha \beta} \big(\alpha^2+\beta(\beta-1)-\alpha(4\beta+1)\big)  \Big)
+\frac{{\cal F}_{\alpha \beta}^3 m_Q^2 (\alpha + \beta -1)^4 }{3^2\times5\times2^{20}\alpha^6\beta^6} \nonumber\\
&\times&\bigg ( -4 m_Q^2 \Big(\alpha^4 +\alpha^3(\beta-1) +\alpha\beta^3+\beta^3(\beta-1)\Big)+{\cal F}_{\alpha \beta}\Big(7 \alpha^3 -3 \alpha^2 (\beta-1) \nonumber\\
&-&3\alpha \beta^2 +\beta^2(7\beta+3) \Big) \bigg) \bigg{\}}\;,\\
\rho^{\langle \bar{q} q \rangle^2}(s)&=&- \frac{\langle \bar{q} q \rangle^2}{2^{10}\pi^6} \int^{\alpha_{max}}_{\alpha_{min}} d \alpha \int^{1 - \alpha}_{\beta_{min}} d \beta \frac{2 m_Q^2 {\cal F}_{\alpha \beta}^3 (\alpha+\beta-1)^2+{\cal F}_{\alpha \beta}^4 (\alpha+\beta - 1)}{3\alpha^3\beta^3}\;,\\
\rho^{\langle G^3 \rangle}(s) &=&\frac{\langle G^3 \rangle}{5\times3^2\times2^{21}\pi^{10}}\int_{\alpha_{min}}^{\alpha_{max}} d \alpha \int_{\beta_{min}}^{1-\alpha} d \beta \frac{ {\cal F}_{\alpha \beta}^2  (\alpha +\beta -1)^4}{\alpha^6\beta^6}  \nonumber\\
&\times& \bigg( 5(\alpha^3+\beta^3) {\cal F}_{\alpha \beta}^2+4 {\cal F}_{\alpha \beta} m_Q^2 \Big( 13\alpha^4+3\alpha^3(\beta-1)+3\alpha\beta^3\nonumber\\
&+&\beta^3(13\beta-3) \Big)+12  m_Q^4 \Big(\alpha^5+\alpha^4(\beta-1)+\alpha\beta^4+\beta^4(\beta-1) \Big)\bigg)\; ,\\
\rho^{\langle \bar{q} q \rangle \langle \bar{q} G q \rangle}(s)&=&\frac{-{\cal N}_i\langle \bar{q} q \rangle \langle \bar{q} G q \rangle}{2^9\pi^6}\int_{\alpha_{min}}^{\alpha_{max}} d \alpha \int_{\beta_{min}}^{1-\alpha} d \beta \frac{{\cal F}_{\alpha \beta}^{3}+3 m_Q^{2}{\cal F}_{\alpha \beta}^2(\alpha+\beta-1)}{ \alpha^{2} \beta^{2}}\; ,\\
\rho^{\langle \bar{q} G q \rangle^2}&=&\frac{{\cal N}_i\langle \bar{q} G q \rangle^2}{2^{11}\pi^6}\int_{\alpha_{min}}^{\alpha_{max}} d \alpha \bigg{\{} \frac{3{\cal H}_\alpha^{2} }{2 (1-\alpha) \alpha}+\int_{\beta_{min}}^{1-\alpha} d \beta \frac{3{\cal F}_{\alpha \beta} m_Q^{2} }{ \alpha \beta}\bigg{\}} \; ,\\
\rho^{\langle \bar{q} q\rangle^4}&=&\int_{\alpha_{min}}^{\alpha_{max}} d \alpha \frac{{\cal H}_\alpha-m_Q^2}{24 \pi^{2}}\langle \bar{q} q\rangle^4\;.
\end{eqnarray}

\subsection{The spectral densities for $0^{+-}$ baryonium state in Eqs.~(\ref{Ja0+-}) and (\ref{Jb0+-})  }
\begin{eqnarray}
\rho^{pert}(s)&=&\int_{\alpha_{min}}^{\alpha_{max}} d \alpha \int_{\beta_{min}}^{1-\alpha} d \beta \bigg{\{}\frac{{\cal F}_{\alpha \beta}^{6}(\alpha+\beta-1)^{4}(5{\cal F}_{\alpha \beta}+7m_Q^2(\alpha+\beta-1)}{3\times7\times5^2\times2^{19}\pi^{10} \alpha^{6} \beta^{6}}\bigg{\}}\; ,\\
\rho^{\langle G^2 \rangle}(s)&=&\frac{\langle G^2 \rangle}{\pi^{10}}\int_{\alpha_{min}}^{\alpha_{max}} d \alpha \int_{\beta_{min}}^{1-\alpha} d \beta \bigg{\{}{\cal N}_i \frac{{\cal F}_{\alpha \beta}^{4} (\alpha+\beta-1)^{2} (5 m_Q^{2}(\alpha+\beta-1)+3{\cal F}_{\alpha \beta})}{3\times5\times2^{19} \alpha^{4} \beta^{4}}\nonumber\\
&+&\frac{{\cal F}_{\alpha \beta}^{3}m_Q^{2}(\alpha+\beta-1)^{4}}{5\times3^2\times2^{21} \alpha^{6} \beta^{6}}\left(4 m_Q^{2}\left(\alpha^{4}+\alpha^{3}(\beta-1)+\alpha \beta^{3}+(\beta-1) \beta^{3}\right)\right. \nonumber\\
&+&{\cal F}_{\alpha \beta}\left(2 \alpha^{3}-3 \alpha^{2}(\beta-1)-3 \alpha \beta^{2}+\beta^{2}(3+2 \beta)) \right)\bigg{\}}\; ,\\
\rho^{\langle G^3 \rangle}(s) &=&\frac{\langle G^3 \rangle}{5\times3^2\times2^{23}\pi^{10}}\int_{\alpha_{min}}^{\alpha_{max}} \frac{d \alpha}{ \alpha ^6} \int_{\beta_{min}}^{1-\alpha} \frac{d \beta}{ \beta ^6} {\cal F}_{\alpha \beta}^2  (\alpha +\beta -1)^4  \nonumber\\
&\times&\bigg{\{}5 {\cal F}_{\alpha \beta}^2(\alpha^3+\beta^3)+8 {\cal F}_{\alpha \beta} m_Q^2 (8 \alpha^4 +3\alpha^3(\beta-1)+3\alpha \beta^3 \nonumber\\
&-&3\beta^3+8\beta^4)+24 \alpha  m_Q^4 (\alpha +\beta -1)(\alpha^4+\beta^4)\bigg{\}}\; ,\\
\rho^{\langle \bar{q} q\rangle^4}&=&-\int_{\alpha_{min}}^{\alpha_{max}} d \alpha \frac{3{\cal H}_\alpha-2m_Q^2}{144 \pi^{2}}\langle \bar{q} q\rangle^4\;,
\end{eqnarray}
where ${\cal N}_A=1$ and ${\cal N}_B=0$.

\subsection{The spectral densities for $0^{+-}$ baryonium state in Eqs.~(\ref{Jc0+-}) and (\ref{Jd0+-})  }

\begin{eqnarray}
\rho^{pert} (s) &=&  \int^{\alpha_{max}}_{\alpha_{min}} d \alpha \int^{1 - \alpha}_{\beta_{min}} d \beta  \frac{{\cal F}^6_{\alpha \beta} (\alpha + \beta - 1)^4 \Big(7  m_Q^2 (\alpha + \beta - 1)+ 10{\cal F}_{\alpha \beta} \Big) }{3\times 7 \times 5^2 \times 2^{18}\pi^{10} \alpha^6 \beta^6}\;,\\
\rho^{\langle G^2 \rangle}(s) &=& \frac{\langle g_s^2 G^2\rangle}{\pi^{10}} \int^{\alpha_{max}}_{\alpha_{min}} d \alpha \int^{1 - \alpha}_{\beta_{min}} d \beta \bigg{\{}- \frac{{\cal F}_{\alpha \beta}^4 (\alpha + \beta -1)^2}{3\times5\times2^{21}\alpha^5\beta^5}\Big(5m_Q^2(\alpha+\beta-1)^2\nonumber\\
&\times&(\alpha+\beta)+4{\cal F}_{\alpha \beta} \big(\alpha^2+\beta(\beta-1)-\alpha(4\beta+1)\big)  \Big)
+\frac{{\cal F}_{\alpha \beta}^3 m_Q^2 (\alpha + \beta -1)^4 }{3^2\times5\times2^{20}\alpha^6\beta^6} \nonumber\\
&\times&\bigg ( 4 m_Q^2 \Big(\alpha^4 +\alpha^3(\beta-1) +\alpha\beta^3+\beta^3(\beta-1)\Big)+{\cal F}_{\alpha \beta}\Big(13 \alpha^3 +3 \alpha^2 (\beta-1) \nonumber\\
&+&3\alpha \beta^2 +\beta^2(13\beta-3) \Big) \bigg) \bigg{\}}\;,\\
\rho^{\langle \bar{q} q \rangle^2}(s)&=&\frac{{\cal N}_i \langle \bar{q} q \rangle^2}{3\times2^{10}\pi^6} \int^{\alpha_{max}}_{\alpha_{min}} d \alpha \int^{1 - \alpha}_{\beta_{min}} d \beta \frac{2 m_Q^2 {\cal F}_{\alpha \beta}^3 (\alpha+\beta-1)^2-{\cal F}_{\alpha \beta}^4 (\alpha+\beta - 1)}{\alpha^3\beta^3}\;,\\
\rho^{\langle G^3 \rangle}(s) &=&\frac{\langle G^3 \rangle}{5\times3^2\times2^{21}\pi^{10}}\int_{\alpha_{min}}^{\alpha_{max}} d \alpha \int_{\beta_{min}}^{1-\alpha} d \beta \frac{ {\cal F}_{\alpha \beta}^2  (\alpha +\beta -1)^4}{\alpha^6\beta^6}  \nonumber\\
&\times& \bigg( 5(\alpha^3+\beta^3) {\cal F}_{\alpha \beta}^2+4 {\cal F}_{\alpha \beta} m_Q^2 \Big( 7\alpha^4-3\alpha^3(\beta-1)-3\alpha\beta^3\nonumber\\
&+&\beta^3(7\beta+3) \Big)-12  m_Q^4 \Big(\alpha^5+\alpha^4(\beta-1)+\alpha\beta^4+\beta^4(\beta-1) \Big)\bigg)\; ,\\
\rho^{\langle \bar{q} q \rangle \langle \bar{q} G q \rangle}(s)&=&\frac{\langle \bar{q} q \rangle \langle \bar{q} G q \rangle}{3\times2^9\pi^6}\int_{\alpha_{min}}^{\alpha_{max}} d \alpha \int_{\beta_{min}}^{1-\alpha} d \beta \frac{{\cal F}_{\alpha \beta}^{3}+3 m_Q^{2}{\cal F}_{\alpha \beta}^2(\alpha+\beta-1)}{ \alpha^{2} \beta^{2}}\; ,\\
\rho^{\langle \bar{q} G q \rangle^2}&=&\frac{\langle \bar{q} G q \rangle^2}{2^{11}\pi^6}\int_{\alpha_{min}}^{\alpha_{max}} d \alpha \bigg{\{} \frac{{\cal H}_\alpha^{2} }{2 (\alpha-1) \alpha}-\int_{\beta_{min}}^{1-\alpha} d \beta \frac{{\cal F}_{\alpha \beta} m_Q^{2} }{ \alpha \beta}\bigg{\}} \; ,\\
\rho^{\langle \bar{q} q\rangle^4}&=&-\int_{\alpha_{min}}^{\alpha_{max}} d \alpha \frac{m_Q^2+3{\cal H}_\alpha}{72 \pi^{2}}\langle \bar{q} q\rangle^4\;,
\end{eqnarray}
where ${\cal N}_C=1$ and ${\cal N}_D=-3$.

\subsection{The spectral densities for $0^{+-}$ baryonium state in Eqs.~(\ref{Je0+-}) and (\ref{Jf0+-})  }

\begin{eqnarray}
\rho^{pert} (s) &=&  \int^{\alpha_{max}}_{\alpha_{min}} d \alpha \int^{1 - \alpha}_{\beta_{min}} d \beta  \frac{{\cal F}^6_{\alpha \beta} (\alpha + \beta - 1)^4 \Big(10{\cal F}_{\alpha \beta}-7  m_Q^2 (\alpha + \beta - 1) \Big) }{3\times 7 \times 5^2 \times 2^{18}\pi^{10} \alpha^6 \beta^6}\;,\\
\rho^{\langle G^2 \rangle}(s) &=& \frac{\langle g_s^2 G^2\rangle}{\pi^{10}} \int^{\alpha_{max}}_{\alpha_{min}} d \alpha \int^{1 - \alpha}_{\beta_{min}} d \beta \bigg{\{} \frac{{\cal F}_{\alpha \beta}^4 (\alpha + \beta -1)^2}{3\times5\times2^{21}\alpha^5\beta^5}\Big(5m_Q^2(\alpha+\beta-1)^2\nonumber\\
&\times&(\alpha+\beta)-4{\cal F}_{\alpha \beta} \big(\alpha^2+\beta(\beta-1)-\alpha(4\beta+1)\big)  \Big)
+\frac{{\cal F}_{\alpha \beta}^3 m_Q^2 (\alpha + \beta -1)^4 }{3^2\times5\times2^{20}\alpha^6\beta^6} \nonumber\\
&\times&\bigg (- 4 m_Q^2 \Big(\alpha^4 +\alpha^3(\beta-1) +\alpha\beta^3+\beta^3(\beta-1)\Big)+{\cal F}_{\alpha \beta}\Big(7 \alpha^3 -3 \alpha^2 (\beta-1) \nonumber\\
&-&3\alpha \beta^2 +\beta^2(7\beta+3) \Big) \bigg) \bigg{\}}\;,\\
\rho^{\langle \bar{q} q \rangle^2}(s)&=&\frac{{\cal N}_i \langle \bar{q} q \rangle^2}{2^{10}\pi^6} \int^{\alpha_{max}}_{\alpha_{min}} d \alpha \int^{1 - \alpha}_{\beta_{min}} d \beta \frac{ {{\cal F}_{\alpha \beta}^4 (\alpha+\beta - 1)-2 m_Q^2\cal F}_{\alpha \beta}^3 (\alpha+\beta-1)^2}{\alpha^3\beta^3}\;,\\
\rho^{\langle G^3 \rangle}(s) &=&\frac{\langle G^3 \rangle}{5\times3^2\times2^{21}\pi^{10}}\int_{\alpha_{min}}^{\alpha_{max}} d \alpha \int_{\beta_{min}}^{1-\alpha} d \beta \frac{ {\cal F}_{\alpha \beta}^2  (\alpha +\beta -1)^4}{\alpha^6\beta^6}  \nonumber\\
&\times& \bigg( 5(\alpha^3+\beta^3) {\cal F}_{\alpha \beta}^2+4 {\cal F}_{\alpha \beta} m_Q^2 \Big( 7\alpha^4-3\alpha^3(\beta-1)-3\alpha\beta^3\nonumber\\
&+&\beta^3(7\beta+3) \Big)-12  m_Q^4 \Big(\alpha^5+\alpha^4(\beta-1)+\alpha\beta^4+\beta^4(\beta-1) \Big)\bigg)\; ,\\
\rho^{\langle \bar{q} q \rangle \langle \bar{q} G q \rangle}(s)&=&\frac{{\cal N}_i\langle \bar{q} q \rangle \langle \bar{q} G q \rangle}{2^9\pi^6}\int_{\alpha_{min}}^{\alpha_{max}} d \alpha \int_{\beta_{min}}^{1-\alpha} d \beta \frac{-{\cal F}_{\alpha \beta}^{3}+3 m_Q^{2}{\cal F}_{\alpha \beta}^2(\alpha+\beta-1)}{ \alpha^{2} \beta^{2}}\; ,\\
\rho^{\langle \bar{q} G q \rangle^2}&=&\frac{-3{\cal N}_i\langle \bar{q} G q \rangle^2}{2^{11}\pi^6}\int_{\alpha_{min}}^{\alpha_{max}} d \alpha \bigg{\{} \frac{{\cal H}_\alpha^{2} }{2 (\alpha-1) \alpha}+\int_{\beta_{min}}^{1-\alpha} d \beta \frac{{\cal F}_{\alpha \beta} m_Q^{2} }{ \alpha \beta}\bigg{\}} \; ,\\
\rho^{\langle \bar{q} q\rangle^4}&=&\int_{\alpha_{min}}^{\alpha_{max}} d \alpha \frac{m_Q^2+3{\cal H}_\alpha}{72 \pi^{2}}\langle \bar{q} q\rangle^4\;,
\end{eqnarray}
where ${\cal N}_C=1$ and ${\cal N}_D=-1$.

\end{widetext}

\begin{thebibliography}{99}

\bibitem{Choi:2003ue}
  S.~K.~Choi {\it et al.} [Belle Collaboration],
  %``Observation of a narrow charmonium - like state in exclusive B+- ---> K+- pi+ pi- J / psi decays,''
  Phys.\ Rev.\ Lett.\  {\bf 91}, 262001 (2003).
 %

%
\bibitem{GellMann:1964nj}
  M.~Gell-Mann,
  %``A Schematic Model of Baryons and Mesons,''
  Phys.\ Lett.\  {\bf 8}, 214 (1964).

\bibitem{Zweig}
  G.~Zweig, Report No. CERN-TH-401.
  
  %other investigations on exotic states
 
 %\cite{Jiao:2009ra}
\bibitem{Jiao:2009ra}
C.~K.~Jiao, W.~Chen, H.~X.~Chen and S.~L.~Zhu,
%``The Possible J**PC = 0-- Exotic State,''
Phys. Rev. D \textbf{79}, 114034 (2009).
%doi:10.1103/PhysRevD.79.114034
%[arXiv:0905.0774 [hep-ph]].
%20 citations counted in INSPIRE as of 16 Oct 2022
 
 %\cite{LEE:2020eif}
\bibitem{LEE:2020eif}
H.~J.~LEE,
%``Discussion on the $J^{PC}=0^{--}$ Tetraquark of $u$ and $d$ Quarks within the QCD sum rule,''
New Phys. Sae Mulli \textbf{70}, 836 (2020).
%doi:10.3938/NPSM.70.836
%2 citations counted in INSPIRE as of 16 Oct 2022
 
 
 %\cite{Shen:2010ky}
\bibitem{Shen:2010ky}
L.~L.~Shen, X.~L.~Chen, Z.~G.~Luo, P.~Z.~Huang, S.~L.~Zhu, P.~F.~Yu and X.~Liu,
%``The Molecular systems composed of the charmed mesons in the $H\bar{S}+h.c.$ doublet,''
Eur. Phys. J. C \textbf{70}, 183 (2010).
%doi:10.1140/epjc/s10052-010-1441-0
%[arXiv:1005.0994 [hep-ph]].
%23 citations counted in INSPIRE as of 16 Oct 2022

%\cite{Wang:2021lkg}
\bibitem{Wang:2021lkg}
Z.~G.~Wang and Q.~Xin,
%``Analysis of the pseudoscalar hidden-charm tetraquark states with the QCD sum rules,''
Nucl. Phys. B \textbf{978}, 115761 (2022).
%doi:10.1016/j.nuclphysb.2022.115761
%[arXiv:2112.04776 [hep-ph]].
%3 citations counted in INSPIRE as of 16 Oct 2022

%\cite{General:2007bk}
\bibitem{General:2007bk}
I.~J.~General, P.~Wang, S.~R.~Cotanch and F.~J.~Llanes-Estrada,
%``Light 1-+ exotics: Molecular resonances,''
Phys. Lett. B \textbf{653}, 216 (2007).
%doi:10.1016/j.physletb.2007.08.015
%[arXiv:0707.1286 [hep-ph]].
%19 citations counted in INSPIRE as of 16 Oct 2022

%\cite{Wan:2022xkx}
\bibitem{Wan:2022xkx}
B.~D.~Wan, S.~Q.~Zhang and C.~F.~Qiao,
Phys. Rev. D \textbf{106}, 074003 (2022).
%doi:10.1103/PhysRevD.106.074003
%[arXiv:2203.14014 [hep-ph]].
%9 citations counted in INSPIRE as of 16 Oct 2022

%\cite{Huang:2016rro}
\bibitem{Huang:2016rro}
Z.~R.~Huang, W.~Chen, T.~G.~Steele, Z.~F.~Zhang and H.~Y.~Jin,
%``Investigation of the light four-quark states with exotic $J^{PC}=0^{--}$,''
Phys. Rev. D \textbf{95}, 076017 (2017).
%doi:10.1103/PhysRevD.95.076017
%[arXiv:1610.02081 [hep-ph]].
%14 citations counted in INSPIRE as of 16 Oct 2022

%\cite{Liu:2005rc}
\bibitem{Liu:2005rc}
Y.~Liu and X.~Q.~Luo,
%``Estimate of the charmed 0-- hybrid meson spectrum from quenched lattice QCD,''
Phys. Rev. D \textbf{73}, 054510 (2006).
%doi:10.1103/PhysRevD.73.054510
%[arXiv:hep-lat/0511015 [hep-lat]].
%29 citations counted in INSPIRE as of 16 Oct 2022

%\cite{Chetyrkin:2000tj}
\bibitem{Chetyrkin:2000tj}
K.~G.~Chetyrkin and S.~Narison,
%``Light hybrid mesons in QCD,''
Phys. Lett. B \textbf{485}, 145 (2000).
%doi:10.1016/S0370-2693(00)00621-3
%[arXiv:hep-ph/0003151 [hep-ph]].
%65 citations counted in INSPIRE as of 16 Oct 2022

%\cite{Govaerts:1984bk}
\bibitem{Govaerts:1984bk}
J.~Govaerts, F.~de Viron, D.~Gusbin and J.~Weyers,
%``{QCD} Sum Rules and Hybrid Mesons,''
Nucl. Phys. B \textbf{248}, 1 (1984).
%doi:10.1016/0550-3213(84)90583-2
%101 citations counted in INSPIRE as of 16 Oct 2022

%\cite{General:2006ed}
\bibitem{General:2006ed}
I.~J.~General, S.~R.~Cotanch and F.~J.~Llanes-Estrada,
%``QCD Coulomb gauge approach to hybrid mesons,''
Eur. Phys. J. C \textbf{51}, 347 (2007).
%doi:10.1140/epjc/s10052-007-0298-3
%[arXiv:hep-ph/0609115 [hep-ph]].
%49 citations counted in INSPIRE as of 16 Oct 2022

%\cite{Ishida:1991mx}
\bibitem{Ishida:1991mx}
S.~Ishida, H.~Sawazaki, M.~Oda and K.~Yamada,
%``Decay properties of hybrid mesons with a massive constituent gluon and search for their candidates,''
Phys. Rev. D \textbf{47}, 179 (1993).
%5doi:10.1103/PhysRevD.47.179
%40 citations counted in INSPIRE as of 16 Oct 2022

%\cite{Qiao:2014vva}
\bibitem{Qiao:2014vva}
C.~F.~Qiao and L.~Tang,
%``Finding the $0^{--}$ Glueball,''
Phys. Rev. Lett. \textbf{113}, 221601 (2014).
%doi:10.1103/PhysRevLett.113.221601
%[arXiv:1408.3995 [hep-ph]].
%28 citations counted in INSPIRE as of 16 Oct 2022

%\cite{Tang:2015twt}
\bibitem{Tang:2015twt}
L.~Tang and C.~F.~Qiao,
%``Mass spectra of $0^{+-}$, $1^{-+}$, and $2^{+-}$ exotic glueballs,''
Nucl. Phys. B \textbf{904}, 282 (2016).
%doi:10.1016/j.nuclphysb.2016.01.017
%[arXiv:1509.00305 [hep-ph]].
%17 citations counted in INSPIRE as of 16 Oct 2022

%\cite{Cotanch:2006wv}
\bibitem{Cotanch:2006wv}
S.~R.~Cotanch, I.~J.~General and P.~Wang,
%``QCD Coulomb Gauge Approach to Exotic Hadrons,''
Eur. Phys. J. A \textbf{31}, 656 (2007).
%doi:10.1140/epja/i2006-10234-2
%[arXiv:hep-ph/0610071 [hep-ph]].
%11 citations counted in INSPIRE as of 16 Oct 2022

%\cite{Bellantuono:2015fia}
\bibitem{Bellantuono:2015fia}
L.~Bellantuono, P.~Colangelo and F.~Giannuzzi,
%``Holographic Oddballs,''
JHEP \textbf{10}, 137 (2015).
%doi:10.1007/JHEP10(2015)137
%[arXiv:1507.07768 [hep-ph]].
%19 citations counted in INSPIRE as of 16 Oct 2022

%\cite{Chen:2021cjr}
\bibitem{Chen:2021cjr}
H.~X.~Chen, W.~Chen and S.~L.~Zhu,
%``Toward the existence of the odderon as a three-gluon bound state,''
Phys. Rev. D \textbf{103}, L091503 (2021).
%doi:10.1103/PhysRevD.103.L091503
%[arXiv:2103.17201 [hep-ph]].
%9 citations counted in INSPIRE as of 16 Oct 2022

%\cite{Zhang:2021itx}
\bibitem{Zhang:2021itx}
L.~Zhang, C.~Chen, Y.~Chen and M.~Huang,
%``Spectra of glueballs and oddballs and the equation of state from holographic QCD,''
Phys. Rev. D \textbf{105}, 026020 (2022).
%doi:10.1103/PhysRevD.105.026020
%[arXiv:2106.10748 [hep-ph]].
%9 citations counted in INSPIRE as of 16 Oct 2022

%\cite{Zhang:2022obn}
\bibitem{Zhang:2022obn}
S.~Q.~Zhang, B.~D.~Wan, L.~Tang and C.~F.~Qiao,
%``Gluonic nature of the newly observed state X(2600),''
Phys. Rev. D \textbf{106}, 074010 (2022).
%doi:10.1103/PhysRevD.106.074010
%[arXiv:2206.13133 [hep-ph]].
%1 citations counted in INSPIRE as of 22 Apr 2024

%\cite{Chen:2021bck}
\bibitem{Chen:2021bck}
H.~X.~Chen, W.~Chen and S.~L.~Zhu,
%``Two- and three-gluon glueballs of C=+,''
Phys. Rev. D \textbf{104}, 094050 (2021).
%doi:10.1103/PhysRevD.104.094050
%[arXiv:2107.05271 [hep-ph]].
%21 citations counted in INSPIRE as of 22 Apr 2024

%\cite{Ballon-Bayona:2017sxa}
\bibitem{Ballon-Bayona:2017sxa}
A.~Ballon-Bayona, H.~Boschi-Filho, L.~A.~H.~Mamani, A.~S.~Miranda and V.~T.~Zanchin,
%``Effective holographic models for QCD: glueball spectrum and trace anomaly,''
Phys. Rev. D \textbf{97}, 046001 (2018).
%doi:10.1103/PhysRevD.97.046001
%[arXiv:1708.08968 [hep-th]].
%37 citations counted in INSPIRE as of 22 Apr 2024
 
 %end other investigation


%exotic states in experiments

 
%\cite{BESIII:2022riz}
\bibitem{BESIII:2022riz}
M.~Ablikim \textit{et al.} [BESIII],
%``Observation of an Isoscalar Resonance with Exotic JPC=1-+ Quantum Numbers in J/\ensuremath{\psi}\textrightarrow{}\ensuremath{\gamma}\ensuremath{\eta}\ensuremath{\eta}',''
Phys. Rev. Lett. \textbf{129}, 192002 (2022).
%doi:10.1103/PhysRevLett.129.192002
%[arXiv:2202.00621 [hep-ex]].
%28 citations counted in INSPIRE as of 21 Nov 2022
 
%\cite{BESIII:2022iwi}
\bibitem{BESIII:2022iwi}
M.~Ablikim \textit{et al.} [BESIII],
%``Partial wave analysis of J/\ensuremath{\psi}\textrightarrow{}\ensuremath{\gamma}\ensuremath{\eta}\ensuremath{\eta}',''
Phys. Rev. D \textbf{106}, 072012 (2022).
%doi:10.1103/PhysRevD.106.072012
%[arXiv:2202.00623 [hep-ex]].
%19 citations counted in INSPIRE as of 21 Nov 2022


 %end exotic states in experiments

%deuteron
\bibitem{Weinberg:1962hj}
  S.~Weinberg,
  Phys.\ Rev.\  {\bf 130}, 776 (1963);
  Phys.\ Rev.\  {\bf 131}, 440 (1963);
  Phys.\ Rev.\  {\bf 137}, B672 (1965).
%end deuteron


\bibitem{Fermi:1949voc}
E.~Fermi and C.~N.~Yang,
%``ARE MESONS ELEMENTARY PARTICLES?,''
Phys.\ Rev. {\bf 76}, 1739 (1949).
%doi:10.1103/PhysRev.76.1739
%251 citations counted in INSPIRE as of 31 Jul 2021

% heavy baryonium

\bibitem{Qiao:2005av}
  C.~F.~Qiao,
  %``One explanation for the exotic state Y(4260),''
  Phys.\ Lett.\ B {\bf 639}, 263 (2006).

\bibitem{Qiao:2007ce}
  C.~F.~Qiao,
  %``A Uniform description of the states recently observed at B-factories,''
  J.\ Phys.\ G {\bf 35}, 075008 (2008).

\bibitem{Chen:2011cta}
  Y.~D.~Chen and C.~F.~Qiao,
  %``Baryonium Study in Heavy Baryon Chiral Perturbation Theory,''
  Phys.\ Rev.\ D {\bf 85}, 034034 (2012).

\bibitem{Chen:2013sba}
  Y.~D.~Chen, C.~F.~Qiao, P.~N.~Shen and Z.~Q.~Zeng,
  %``Revisiting the spectrum of baryonium in heavy baryon chiral perturbation theory,''
  Phys.\ Rev.\ D {\bf 88}, 114007 (2013).

  \bibitem{Wan:2019ake}
B.~D.~Wan, L.~Tang and C.~F.~Qiao,
%``Hidden-bottom and -charm hexaquark states in QCD sum rules,''
Eur.\ Phys.\ J.\ C {\bf 80}, 121 (2020).
%doi:10.1140/epjc/s10052-020-7701-8
%[arXiv:1912.12046 [hep-ph]].
%4 citations counted in INSPIRE as of 13 Jul 2021

\bibitem{Chen:2016ymy}
  H.~X.~Chen, D.~Zhou, W.~Chen, X.~Liu and S.~L.~Zhu,
  %``Searching for hidden-charm baryonium signals in QCD sum rules,''
  Eur.\ Phys.\ J.\ C {\bf 76}, 602 (2016).

\bibitem{Liu:2007tj}
C.~Liu,
%``Baryonium, tetra-quark state and glue-ball in large N(c) QCD,''
Eur.\ Phys.\ J.\ C {\bf 53}, 413 (2008).
%doi:10.1140/epjc/s10052-007-0471-8
%[arXiv:0710.4185 [hep-ph]].
%27 citations counted in INSPIRE as of 13 Jul 2021

%\cite{Wang:2021qmn}
\bibitem{Wang:2021qmn}
X.~W.~Wang, Z.~G.~Wang and G.~l.~Yu,
%``Study of $\Lambda _c\Lambda _c$ dibaryon and $\Lambda _c{\bar{\Lambda }}_c$ baryonium states via QCD sum rules,''
Eur. Phys. J. A \textbf{57}, 275 (2021).
%doi:10.1140/epja/s10050-021-00576-8
%[arXiv:2107.04751 [hep-ph]].
%10 citations counted in INSPIRE as of 10 Oct 2022

%\cite{Wan:2022uie}
\bibitem{Wan:2022uie}
B.~D.~Wan and C.~F.~Qiao,
%``The tetra-heavy baryonium spectra,''
[arXiv:2208.14042 [hep-ph]].
%0 citations counted in INSPIRE as of 10 Oct 2022
 
 %\cite{Liu:2021gva}
\bibitem{Liu:2021gva}
Z.~Liu, H.~T.~An, Z.~W.~Liu and X.~Liu,
%``Where are the hidden-charm hexaquarks?,''
Phys. Rev. D \textbf{105}, 034006 (2022).
%doi:10.1103/PhysRevD.105.034006
%[arXiv:2112.02510 [hep-ph]].
%6 citations counted in INSPIRE as of 10 Oct 2022

%\cite{Wan:2021vny}
\bibitem{Wan:2021vny}
B.~D.~Wan, S.~Q.~Zhang and C.~F.~Qiao,
%``Light baryonium spectrum,''
Phys.\ Rev.\ D {\bf 105}, 014016 (2022).
%doi:10.1103/PhysRevD.105.014016
%[arXiv:2109.07130 [hep-ph]].
%1 citations counted in INSPIRE as of 15 Mar 2022
  
%\cite{BESIII:2022tvj}
\bibitem{BESIII:2022tvj}
M.~Ablikim \textit{et al.} [BESIII],
%``Measurement of $e^+e^-\rightarrow\Lambda\bar{\Lambda}\eta$ from 3.5106 to 4.6988 GeV and study of $\Lambda\bar{\Lambda}$ mass threshold enhancement,''
Phys. Rev. D \textbf{107}, 112001 (2023).
%doi:10.1103/PhysRevD.107.112001
%[arXiv:2211.10755 [hep-ex]].
%3 citations counted in INSPIRE as of 22 Nov 2023


\bibitem{Shifman}
  M.A. Shifman, A.I. Vainshtein and V.I. Zakharov,
  Nucl. Phys. {\bf B147}, 385 (1979); ibid, Nucl. Phys. {\bf B147},
  448 (1979).

\bibitem{Albuquerque:2013ija}
  R.~M.~Albuquerque,
  arXiv:1306.4671 [hep-ph].

\bibitem{Tang:2019nwv}
L.~Tang, B.~D.~Wan, K.~Maltman and C.~F.~Qiao,
%``Doubly Heavy Tetraquarks in QCD Sum Rules,''
Phys.\ Rev.\ D {\bf 101}, 094032 (2020).
%doi:10.1103/PhysRevD.101.094032
%[arXiv:1911.10951 [hep-ph]].
%8 citations counted in INSPIRE as of 15 Jul 2021

\bibitem{Matheus:2006xi}
  R.~D'E.~Matheus, S.~Narison, M.~Nielsen and J.~M.~Richard,
  Phys.\ Rev.\ D {\bf 75}, 014005 (2007).

\bibitem{Cui:2011fj}
  C.~Y.~Cui, Y.~L.~Liu and M.~Q.~Huang,
  Phys.\ Rev.\ D {\bf 85}, 074014 (2012).

\bibitem{Narison:2002pw}
  S.~Narison,
  Camb.\ Monogr.\ Part.\ Phys.\ Nucl.\ Phys.\ Cosmol.\  {\bf 17}, 1 (2002).

 \bibitem{P.Col}
  P. Colangelo and A. Khodjamirian, in {\it At the frontier of
  particle physics / Handbook of QCD}, edited by M. Shifman (World
  Scientific, Singapore, 2001), arXiv:hep-ph/0010175.
  
  %\cite{Albuquerque:2016znh}
\bibitem{Albuquerque:2016znh}
R.~Albuquerque, S.~Narison, F.~Fanomezana, A.~Rabemananjara, D.~Rabetiarivony and G.~Randriamanatrika,
%``XYZ-like Spectra from Laplace Sum Rule at N2LO in the Chiral Limit,''
Int. J. Mod. Phys. A \textbf{31}, 1650196 (2016).
%doi:10.1142/S0217751X16501967
%[arXiv:1609.03351 [hep-ph]].
%31 citations counted in INSPIRE as of 22 Apr 2024

\bibitem{Reinders:1984sr}
  L.~J.~Reinders, H.~Rubinstein and S.~Yazaki,
  Phys.\ Rept.\  {\bf 127}, 1 (1985).

\bibitem{Qiao:2013dda}
C.~F.~Qiao and L.~Tang,
%``Interpretation of $Z_c(4025)$ as the hidden charm tetraquark states via QCD Sum Rules,''
Eur.\ Phys.\ J.\ C {\bf 74}, 2810 (2014).
%doi:10.1140/epjc/s10052-014-2810-x
%[arXiv:1308.3439 [hep-ph]].
%46 citations counted in INSPIRE as of 16 Jul 2021

%\cite{Qiao:2013raa}
\bibitem{Qiao:2013raa}
C.~F.~Qiao and L.~Tang,
%``Estimating the mass of the hidden charm $1^+(1^{+})$ tetraquark state via QCD sum rules,''
Eur. Phys. J. C \textbf{74}, 3122 (2014).
%doi:10.1140/epjc/s10052-014-3122-x
%[arXiv:1307.6654 [hep-ph]].
%42 citations counted in INSPIRE as of 19 Mar 2022

\bibitem{Tang:2016pcf}
L.~Tang and C.~F.~Qiao,
%``Tetraquark States with Open Flavors,''
Eur.\ Phys.\ J.\ C {\bf 76}, 558 (2016).
%doi:10.1140/epjc/s10052-016-4436-7
%[arXiv:1603.04761 [hep-ph]].
%36 citations counted in INSPIRE as of 16 Jul 2021

\bibitem{Wan:2020oxt}
B.~D.~Wan and C.~F.~Qiao,
%``About the exotic structure of $Z_{cs}$,''
Nucl.\ Phys.\ B {\bf 968}, 115450 (2021).
%doi:10.1016/j.nuclphysb.2021.115450
%[arXiv:2011.08747 [hep-ph]].
%20 citations counted in INSPIRE as of 15 Jul 2021

 \bibitem{Wan:2020fsk}
B.~D.~Wan and C.~F.~Qiao,
%``Gluonic tetracharm configuration of $X (6900)$,''
Phys.\ Lett.\ B {\bf 817}, 136339 (2021).
%doi:10.1016/j.physletb.2021.136339
%[arXiv:2012.00454 [hep-ph]].
%10 citations counted in INSPIRE as of 15 Jul 2021





\end{thebibliography}
\end{document}